\documentclass[twocolumn,superscriptaddress,letter]{revtex4}%
\usepackage{graphicx}
\usepackage{amsmath}

\providecommand{\U}[1]{\protect \rule{.1in}{.1in}}

\newcommand{\beq}{\begin{eqnarray}}
\newcommand{\eeq}{\end{eqnarray}}
\newcommand{\be}{\begin{equation}}
\newcommand{\ee}{\end{equation}}
\newcommand{\bw}{\begin{widetext}}
\newcommand{\ew}{\end{widetext}}
\newcommand{\ba}{\begin{array}}
\newcommand{\ea}{\end{array}}

\newcommand{\bn}{\mathbf{n}}

\newcommand{\bp}{\mathbf{p}}
\newcommand{\bq}{\mathbf{q}}

\newcommand{\bx}{{\bf x}}
\newcommand{\by}{{\bf y}}

\begin{document}

\title{A Lattice Study of the Two-photon Decay Widths for Scalar and Pseudo-scalar Charmonium}

\author{Ying Chen}
\affiliation{%
Institute of High Energy Physics, Chinese Academy of Sciences, Beijing 100049, China\\
School of Physics, University of Chinese Academy of Sciences, Beijing 100049, China
}%

\author{Ming Gong}
\affiliation{%
Institute of High Energy Physics, Chinese Academy of Sciences, Beijing 100049, China\\
School of Physics, University of Chinese Academy of Sciences, Beijing 100049, China
}%

\author{Ning Li}
\email[Corresponding author. Email: ]{lining@xatu.edu.cn}
\affiliation{%
School of Science, Xi'an Technological University, Xi'an 710032, China
}%

\author{Chuan Liu}%
\affiliation{%
School of Physics and Center for High Energy Physics, Peking
University, Beijing 100871, China\\
Collaborative Innovation Center of Quantum Matter, Beijing 100871, China
}%

\author{Yu-Bin Liu}
\affiliation{%
School of Physics, Nankai University, Tianjin 300071, China
}

\author{Zhaofeng Liu}
\affiliation{%
Institute of High Energy Physics, Chinese Academy of Sciences, Beijing 100049, China\\
School of Physics, University of Chinese Academy of Sciences, Beijing 100049, China
}

\author{Jian-Ping Ma}
\affiliation{%
Institute of Theoretical Physics, Chinese Academy of Sciences, Beijing 100190, China
}

\author{Yu~Meng}
\affiliation{%
School of Physics, Peking University, Beijing 100871, China
}

\author{Chao Xiong}
\affiliation{%
School of Physics, Peking University, Beijing 100871, China
}%

\author{Ke-Long~Zhang}
\affiliation{%
School of Physics, Peking University, Beijing 100871, China
}

 \collaboration{CLQCD Collaboration}

\begin{abstract}
In this exploratory study, two photon decay widths of pseudo-scalar ($\eta_c$) and scalar ($\chi_{c0}$) charmonium
 are computed using two ensembles of $N_f=2$ twisted mass lattice QCD gauge configurations.
 The simulation is performed two lattice ensembles with lattice spacings
$a=0.067$ fm with size $32^3\times{64}$ and $a=0.085$ fm with size $24^3\times{48}$, respectively.
 The results for the decay widths for the two charmonia are obtained which are
 in the right ballpark however smaller than the experimental ones.
 Possible reasons for these discrepancies are discussed.



\end{abstract}


\maketitle

\section{INTRODUCTION}
Charmonium physics plays an important role in the foundation of quantum chromodynamics (QCD)
which is believed to be the fundamental theory for the strong interaction.
Due to its intermediate energy scale and the special features of QCD,
both perturbative and non-perturbative physics show up within
charmonium physics~\cite{Eshraim:2018jkt}. Until now, the best way to study
non-perturbative QCD is lattice QCD, a quantum field theory defined on discrete Euclidean space-time.
Within this formalism, physical quantities are
encoded in various Euclidean correlation functions which in turn can be measured using
Monte Carlo simulations~\cite{Tormo:2013tha,Cornwall:2013zra}.

Recently, two photon decay branching fraction for charmonium has been attracting considerable attentions.
Theoretically, this quantity is considered to provide a probe for the strong coupling constant at the charmonium scale.
It has also been proposed as a sensitive test
of the corrections to the non-relativistic approximation via quark models
or the effective field theories such as non-relativistic QCD (NRQCD).

In addition, considerable progress has been made in recent years in the physics of charmonia via the
investigations from various experimental collaborations, such as, Belle, BaBar, CLEO-c, and BES~\cite{Lees:2010de,Pham:2013ida,Savinov:2013hda,Ablikim:2012xi}.
Generically, there are two ways to measure two-photon branching fraction for charmonium:
one is by reconstructing the charmonium in light hadrons with two-photon fusion at $e^+e^-$ machines,
the other one os from the $p\bar{p}$  annihilation to charmonium.

In this paper, we calculate the two-photon decay width
of the pseudoscalar and scalar charmonium, i.e. $\Gamma(\eta_{c}\rightarrow\gamma\gamma)$
and $\Gamma(\chi_{c0}\rightarrow\gamma\gamma)$,
in lattice QCD  using two ensembles of $N_f=2$ dynamical twisted mass
fermion configurations that were generated by the European Twisted Mass Collaboration(ETMC).
This ensures the so-called automatic $\mathcal{O}(a)$ improvement for on-shell observables
when the twisted mass fermions are at the maximal twist~\cite{Frezzotti:2001ea}.
Lattice computation for the process $\eta_c\rightarrow\gamma\gamma$ has been done
using the same set of configurations in Ref.~\cite{Chen:2016yau}. In this work,
we consider $\eta_c$ and $\chi_{c0}$ simultaneously since they mix with each
other due to lattice artifact of $O(a^2)$.

This paper is organized as follows. In Sect.~\ref{lattice setup},
the calculation strategy for the matrix element for two-photon decay
of charmonium is reviewed which is then related to the double-photon decay rates.
We also outline the strategy for the mass spectrum and
form factor of charmonium. In order to improve
the signals of the corresponding correlation functions, we apply the variation
method to construct the relevant interpolate operators.
In Sect.~\ref{Simulation details}, the simulation details is divided
into several parts:
In Sect.~\ref{simulation configuration}, we give the parameters relevant with the
configurations used in this work.
In Sect.~\ref{Simulation details mass},  mass spectrum of $\eta_c$ and $\chi_{c0}$ are obtained.
In Sect.~\ref{Renormalization factor}, we presents the results of renormalization factor of electromagnetic current operators.
In Sect.~\ref{sec:formfactory}, numerical results of the form factors
are presented which are then converted to the two photon decay width of $\eta_c$  and $\chi_{c0}$.
Final results for the decay widths are presented and compared with previous lattice computations and
the experiments. The discussion and the conclusion can be found in Sect.~\ref{conclusion}.

\section{Strategies for the computation}
\label{lattice setup}

In this section, we briefly review the methods for the calculation of two-photon
decay rate of a charmonium which was presented in Ref.~\cite{Dudek:2006ut}.
According to Lehmann-Symanzik-Zimmermann (LSZ) reduction formula, we can express the
amplitude for two-photon decay of charmonium in the following form:
\bw
\begin{multline}
\label{eq:lsz}
 \langle \gamma(q_1, \iota_1)\gamma(q_2, \iota_2) |M(p_f) \rangle
 = - \lim_{\substack{q'_1 \to q_1 \\ q'_2 \to q_2}} \epsilon^*_\mu(q_1,
 \iota_1) \epsilon^*_\nu(q_2, \iota_2)
 q_1'^2 q_2'^2 \int d^4x d^4y \, e^{i q'_1.y +
 i q'_2.x} \langle \Omega | T\big\{ A^\mu(y) A^\nu(x) \big\} |M(p_f) \rangle.
\end{multline}
\ew
Here $|\Omega\rangle$ designates the QCD vacuum state;
$M$ represents either $\eta_c$ or $\chi_{c0}$ meson state,
depending on which one is calculated, and
$|M(p_f)\rangle$ is the corresponding meson state with four-momentum $p_f$, and $|\gamma(q_i,\iota_i)\rangle$($i=1,2$)
is a single photon state which has the polarization vector $\epsilon(q_i,\iota_i)$
with four-momentum $q_i$ and the helicity $\iota_i$.
Then, treating QED part perturbatively, we can replace the photon field operators by
their corresponding current operators in QCD. We finally arrive at the following equation at the lowest order of QED,
\bw
\beq
\label{eq:big1}
\langle \gamma(q_1, \iota_1)\gamma(q_2, \iota_2) | M(p_f) \rangle
&=&{(-e^2)} \lim_{\substack{q'_1 \to q_1 \\ q'_2 \to q_2}} \epsilon^*_\mu(q_1,
 \iota_1) \epsilon^*_\nu(q_2, \iota_2) q_1'^2 q_2'^2
 \int d^4x d^4y d^4w\ d^4z e^{i q'_1.y +
 i q'_2.x}\, D^{\mu \rho}(y ,z) D^{\nu \sigma}(x, w)\nonumber \\
& &\times\langle
 \Omega | T\big\{ j_\rho(z) j_\sigma(w)\big\} | M(p_f)\rangle
\eeq
\ew
 with $D^{\mu\nu}(y,z)$ being the free photon propagator.
 Basically, each initial/final photon state in the problem is replaced by
 the corresponding electromagnetic current operator that couples to the photon.
 Finally, one needs to compute a three-point function of the form
 $\langle\Omega | T\big\{ j_\rho(z) j_\sigma(w)\big\} | M(p_f)\rangle$ which is
 non-perturbative in nature and can be computed using lattice QCD methods.

Under certain conditions, Eq.~(\ref{eq:big1}) can be analytically
continued from Minkowski space to Euclidean space, yielding
\bw
\beq
\langle\gamma(q_1, \iota_1) \gamma(q_2, \iota_2) |M(p_f)\rangle
&=&
\lim_{t_f-t \to \infty} e^2 \frac{ \epsilon_\mu(q_1,
 \iota_1) \epsilon_\nu(q_2, \iota_2)}{\tfrac{Z_{M}(p_f)}{2 E_{M}(p_f)} e^{-E_{M}(p)(t_f-t)}} \int dt_i
e^{- \omega_1 (t_i -t)}\nonumber\\
& &
\langle \Omega | T\Big\{ \int d^3 \vec{x}\, e^{-i\vec{p_f}.\vec{x}}
\varphi_{M}(\vec{x}, t_f) \int d^3 \vec{y}\, e^{i\vec{q_2}.\vec{y}}
j^\nu(\vec{y}, t) j^\mu(\vec{0}, t_i)\Big\}|\Omega\rangle,  \label{master}
\eeq
\ew
where $\varphi_{M}(\vec{x}, t_f)$ is the field operator for the meson M, $Z_{M}(p_f)$ is the spectral weight factor of two-point function, $\omega_1$ is the energy for the photon at time-slice $t_i$, and $E_{M}(p_f)$ is the energy for
corresponding meson with the momentum $p_f$. Then, the desired amplitude
$\langle\gamma(q_1, \iota_1) \gamma(q_2, \iota_2) |M(p_f)\rangle$
can be obtained once the energies $E_{M}(p_f)$
and the corresponding overlap matrix element
$Z_{M}(p_f)$ are known. These could be obtained from appropriate two-point functions.
In this study, we use the variational method to find the optional interpolation operators to
create/annihinite the $\eta_c$ and $\chi_{c0}$ meson state~\cite{Shultz:2015pfa}.

Generally, an operator $\mathcal{O}^{\dagger}_i$
having definite $J^{PC}$ can produce all QCD eigenstates with the right
quantum numbers
\beq
\mathcal{O}_i^\dagger|\Omega\rangle&=&\sum_n|n\rangle\langle n|\mathcal{O}_i^\dagger|\Omega\rangle.
\eeq
In order to create the desired hadrons from vacuum effectively,
one could employ  a basis of interpolators ${\mathcal{O}_i}$ that share the same quantum numbers
and construct two-point correlation function matrix as follows
\beq
C_{ij}&=&\langle\Omega|\mathcal{O}_i(t)\mathcal{O}_j^\dagger(0)|\Omega\rangle.
\label{eq:TPCF}
\eeq
Here, the operators ${\mathcal{O}_i}$ are color-singlet constructions built
from the basic quark and gluon fields of QCD.
Then, we can express the correlation functions in the following form:
\beq
C_{ij}&=&\sum_n\frac{1}{2E_n}\langle\Omega|\mathcal{O}_i(0)|n\rangle\langle n|\mathcal{O}_j^\dagger(0)|\Omega\rangle
e^{-E_nt}.\nonumber
\eeq
and the optimized interpolators are:
\beq
\label{eq:optimized_op}
\Omega_n^\dagger=\sum_iv_i^n\mathcal{O}_i^\dagger.
\eeq
Therefore, one should obtain the best estimate for the weights $v^n_i$ according to
the solution of the following generalized eigenvalue problem
\beq
C(t)v^n=\lambda_n(t)C(t_0)v^n.
\label{eq:IIIB5}
\eeq
Here, $C(t)$ is the $N\times{N}$ matrix whose elements are the
correlation functions $C_{ij}(t)$ constructed from the basis
of $N$ operators, $v^n$ is a generalized eigenvector.
The generalized eigenvalues, or principal correlators, $\lambda_n(t)$ behave like $e^{-E_n(t-t_0)}$ at large times,
and can be used to determine the spectrum of the states.
In practice, we solve Eq.~(\ref{eq:IIIB5}) independently on each
time-slice $t$, so that for each state $n$, we obtain a time
series of generalized eigenvectors $v^n(t)$. We use $v^n_i$
 chosen on a single time-slice to construct the
optimized operators in Eq.~(\ref{eq:optimized_op}).

 Apart from the two-point functions of $\eta_c$ and $\chi_{c0}$,
 we also need the three-point
 functions $G_{\mu\nu}(t_i,t)$ given by:
 \beq
 G_{\mu\nu}(t_i,t)
 &=&\langle 0 | T\Big\{ \int d^3 \vec{x}\, e^{-i\vec{p_f}.\vec{x}}
\varphi_{M}(\vec{x}, t_f)\nonumber\\
& &\times\int d^3 \vec{y}\, e^{i\vec{q_2}.\vec{y}}
j^\nu(\vec{y}, t) j^\mu(\vec{0}, t_i)\Big\}|0\rangle.
\label{eq:three point1}
 \eeq
We simulate $G_{\mu\nu}(t_i,t)$ on lattices across the temporal
direction while the sink of the meson is fixed.
Then, we repeat this with a varying $t_i$
to integrate over the three point function with an exponential weight $e^{-\omega_1{t_i}}$,
and then to extract the matrix element in Eq.~(\ref{master}).
In particular, we use the optimized interpolators $\Omega_n$ in Eq.~(\ref{eq:optimized_op}) to
create the $\eta_c,(n=1)$ or $\chi_{c0},(n=2)$ state for the field interpolating operator $\varphi_{M}$ in previous formulas.

For the two photon decay of $\eta_c$ meson, the matrix element $\langle\gamma(q_1, \iota_1) \gamma(q_2, \iota_2) |M(p_f)\rangle$
in Eq.~(\ref{master}) can be parameterized using form factor
$F(Q_1^2,Q_2^2)$ as,
\beq
&& \langle\gamma(q_1, \iota_1) \gamma(q_2, \iota_2) |M(p_f)\rangle \nonumber\\
&=&2 (\tfrac{2}{3} e)^2 m_{\eta_c}^{-1} F(Q_1^2,Q_2^2)\epsilon_{\mu\nu\rho\sigma}\epsilon^{\mu}(q_1,\iota_1)\epsilon^{\nu}(q_2,\iota_2)q_1^{\rho}q_2^{\sigma}\nonumber\\
\eeq
where $\epsilon_1, \epsilon_2$ are polarization vectors, $Q_1^2,Q_2^2$ are virtualities and
$q_1, q_2$ are the four-momenta for the two photons.
The corresponding decay width can be expressed in terms of $F(0,0)$,
 \be
\Gamma(\eta_c\rightarrow\gamma\gamma)= \pi \alpha_{em}^2 \tfrac{16}{81}m_{\eta_c} |F(0,0)|^2
\label{eq:decay_width_e}
\ee
with $\alpha_{em}\simeq 1/137$ being the fine structure constant.
Similarly, for $\chi_{c0}$ we have another form factor $G(Q^2_1,Q^2_2)$,
\beq
&&\langle\gamma(q_1, \iota_1) \gamma(q_2, \iota_2) |M(p_f)\rangle \nonumber\\
&=&2 (\tfrac{2}{3} e)^2 m_{\chi_{c0}}^{-1} G(Q_1^2,Q_2^2)(\epsilon_1\cdot\epsilon_2q_1\cdot{q_2}-\epsilon_2\cdot{q_1}\epsilon_1\cdot{q_2})\nonumber\\
\eeq
with the decay width given by
 \beq
\Gamma(\chi_{c0}\rightarrow\gamma\gamma)& =& \pi \alpha_{em}^2 \tfrac{16}{81}m_{\chi_{c0}} |G(0,0)|^2\;.
\label{eq:decay_width_c}
\eeq

\section{Simulation Results}
\label{Simulation details}
\subsection{Simulation setup}
\label{simulation configuration}
In this work, we utilize two ensembles with $N_f=2$ (degenerate $u$ and $d$ quarks) twisted mass configurations.
These configurations are generated by the ETMC at the maximal twist to implement the so-called  automatic
 $\mathcal{O}(a)$ improvement~\cite{Frezzotti:2001ea}.
The explicit parameters for these ensembles are presented in Ref.~\cite{Blossier:2010cr} and the
two ensembles that we utilized are tabulated in Table~\ref{configurations}.
\begin{table}[!htbp]
\begin{center}
\caption{Configuration parameters.}
\begin{tabular}
{cccccccc}\hline\hline
&$\beta$       &$a$[fm]      &$V/{a^4}$            &$a\mu_{sea}$   &$m_{\pi}$[MeV]  &$m_{\pi}L$   &$N_{cfg}$          \\
\hline
Ens.$~B_1$   &$3.9$         &0.085      &$24^3\times{48}$     &0.004           &315              &3.3           &200        \\
Ens.$~C_1$   &$4.05$        &0.067      &$32^3\times{64}$     &0.003           &300              &3.3           &199         \\
\hline\hline
\label{configurations}
\end{tabular}
\end{center}
\end{table}
For the valence sector, we adopt the Osterwalder-Seiler setup which amounts to
introducing two extra twisted doublets for each non-degenerate quark flavors,
namely, $(u,d)$ and $(c,c^{\prime})$ with twisted masses $a\mu_l$ and $a\mu_c$, respectively~\cite{Chen:2011kpa,AbdelRehim:2006ve,Blossier:2007vv,Blossier:2009bx,Frezzotti:2004wz}.
The explicit values of $a\mu_l$ on Ens.$~B_1$ is $0.004$,
and $0.003$ for the Ens.$~C_1$, respectively.
In this simulation, we use the physical mass of $\eta_c$ to set the value of the $a\mu_c$,
and the explicit values are $0.2542$ and $0.2018$ for Ens.$~B_1$ and Ens.$~C_1$ respectively.
In each doublet, the Wilson parameters have opposite signs ($r=-r^{\prime}=1$).
Performing an axial (or chiral) transformation, quark fields in the physical basis
transform into the twisted basis~\cite{Chen:2011kpa}; i.e.,
\beq
\left(
\begin{array}{c}
               u\\
               d
             \end{array}\right)
&=&\exp(i\omega\gamma_5\tau_3/2)\left(\begin{array}{c}
               \chi_u\\
               \chi_d
             \end{array}\right)\nonumber\\
\left(
\begin{array}{c}
               c\\
               c^{\prime}
             \end{array}\right)
&=&\exp(i\omega\gamma_5\tau_3/2)\left(\begin{array}{c}
               \chi_c\\
               \chi_{c^{\prime}}
             \end{array}\right),
\label{traslation}
\eeq
where $\omega$ is the twist angle, and $\omega=\pi/2$ represents maximal twist.
Then, the left of the above equations correspond to quark fields in the physical basis,
and the right correspond to quark fields in the twisted basis.

Before writing out the explicit form of meson operators, one should exploit the symmetry properties of twisted mass LQCD.
We will follow the discussion in reference~\cite{Kalinowski:2015bwa} below.
Isospin $I$ and parity $\mathcal{P}$ are broken by $\mathcal{O}(a^2)$ effects in twisted mass LQCD.
While, a specific combination (i.e. light flavor exchange
combined with parity) is still a symmetry of twisted mass LQCD.
We first write down the interpolating-field operators in the twisted basis,
and build the interpolating operators with the same Wilson
parameters~\cite{Blossier:2009bx}.
For the purpose of $\eta_c$ and $\chi_{c0}$,
we use two basis operators $\mathcal{O}_1(x)=\bar{c}(x)\gamma_5{c}(x)$,  $\mathcal{O}_2=\bar{c}(x){c}(x)$.
According to Eq.~(\ref{traslation}), the two basic operators in twisted basis are given by $\mathcal{O}_1=\bar{\chi}_{c}{\chi_c}$ and $\mathcal{O}_2=\bar{\chi}_{c}\gamma_5{\chi_c}$ which appear to have opposite parity. However, since twisted mass lattice QCD
breaks parity, they in fact mix with each other. Taking into account of this mixing is
 crucial. One needs to go through the solution of the generalized eigenvalue problem
 in Eq.~(\ref{eq:IIIB5}) to obtain the optimized operator that will create the $\eta_c$ and $\chi_{c0}$ meson
 from the vacuum. Without performing this generalized eigenvalue separation,
 it is found that the correct signal of $\chi_{c0}$ cannot be observed in the two-point functions.
 The signal of $\eta_c$ can of course be observed even without considering this mixing effect
 since it is the lightest state under consideration.

\subsection{Mass spectra for $\eta_c$ and $\chi_{c0}$}
\label{Simulation details mass}
\begin{figure*}[!htbp]
\begin{minipage}{0.48\linewidth}
  {\includegraphics[width=9cm]{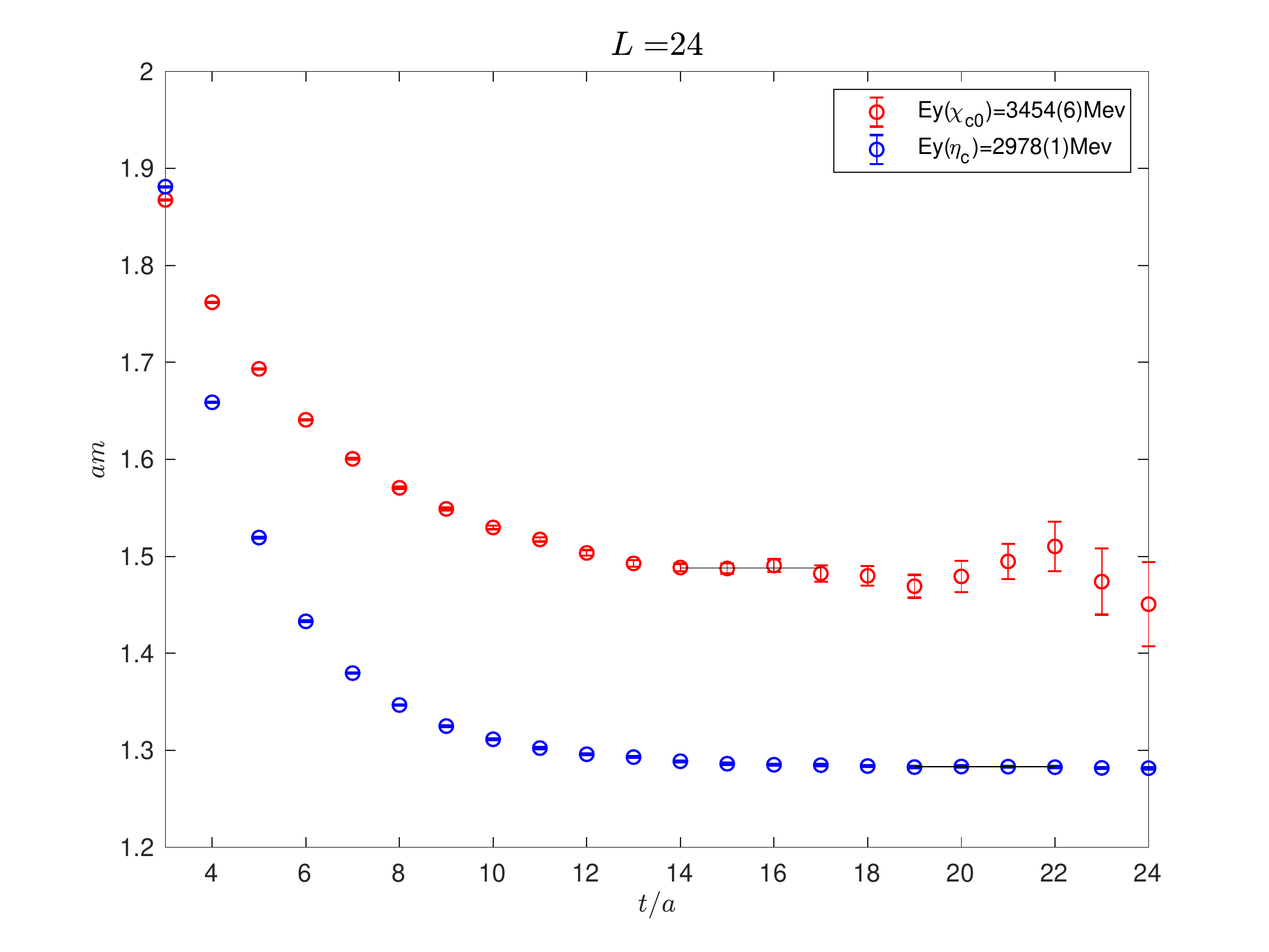}}
  \center (a) Ens.~$B_1$
\end{minipage}
\hfill
\begin{minipage}{0.48\linewidth}
  {\includegraphics[width=9cm]{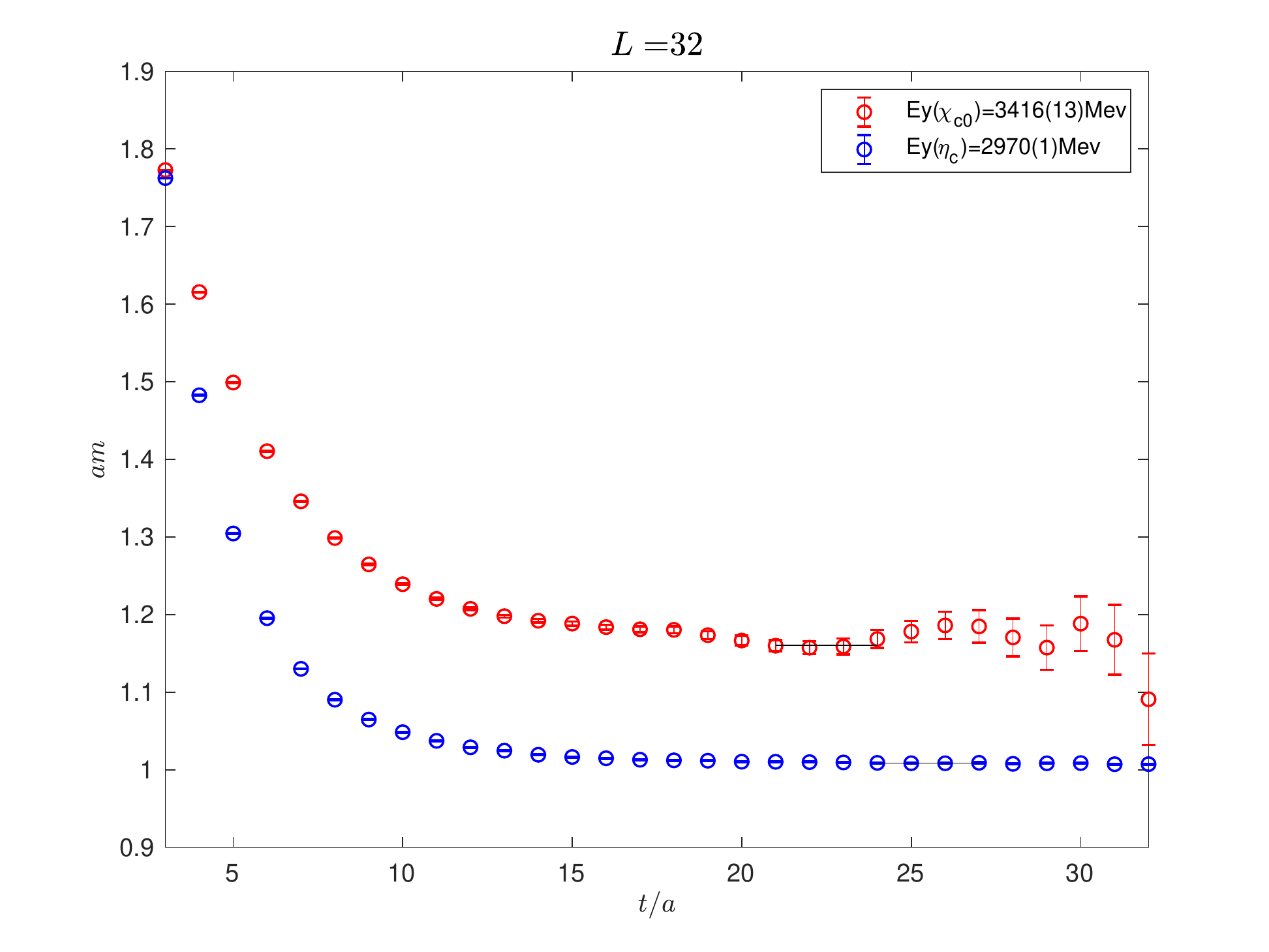}}
  \center (b) Ens.~$C_1$
\end{minipage}
\caption{Mass plateaus of $\eta_c$ (blue point) and $\chi_{c0}$ (red point) for Ens.~{$B_1$} and Ens.~{$C_1$}.
 The horizontal line segments denote the corresponding mass plateaus.}
\label{fig:meson_mass}
\end{figure*}

The eigenvalue $\lambda_n$ in Eq.~(\ref{eq:IIIB5}) corresponding with the
corresponding meson state, i.e. $n=1$ represents $\eta_c$ meson, and $n=2$ represents $\chi_{c0}$ meson .
Since, we use the anti-periodic boundary condition,
 \beq \label{eq:twopoint}
 \lambda_n(t,\bp_f)&\stackrel{t\gg 1}{\longrightarrow}&\frac{|Z_{M}|^2}{E_{M}(\bp_f)}e^{-E_{M}(\bp_f)\cdot\frac
 {T}{2}}\nonumber\\
 & &\times\cosh\left[E_{M}(\bp_f)\cdot\left(\frac T2-t\right)\right]\;.
 \eeq
In practical, we use eigenvalue $\lambda_1$ with $\bp_f=(0,0,0)$ to fit the spectral weight $Z_{\eta_c}$, and
the explicit value of this factor is $0.4416(8)$ and $0.2675(3)$ on Ens.~$B_1$ and Ens.~$C_1$, respectively.
While, we use eigenvalue $\lambda_2$ with $\bp_f=(0,0,0)$ to fit the spectral weight $Z_{\chi_{c0}}$, and
the explicit value of this factor is $0.6699(71)$ and $0.2983(33)$ on Ens.~$B_1$ and Ens.~$C_1$, respectively.
It is easily seen that the mass can also be extracted from,
\beq
\cosh(m_n)&=&\frac{\lambda_n(t-1)+\lambda_n(t+1)}{2\lambda_n(t)},
\label{mass_ralation}
\eeq
with $m_1$ being the mass of $\eta_c$ meson  and $m_2$ being that of $\chi_{c0}$, respectively.
The effective mass plateaus of these mesons for Ens.~{$B_1$} and Ens.~{$C_1$} are illustrated in Fig.~\ref{fig:meson_mass}.
From these mass plateaus, the masses of the mesons are determined and the statistical errors are obtained using jackknife method.
The numerical results for the masses are summarized in Tab.~\ref{tab:mass}.
 Note that the mass values for $\eta_c$ are utilized to fix the valence charm quark mass parameter $a\mu_c$.
 Therefore, only the mass of $\chi_{c0}$ are predictions from this lattice computation.

\begin{table}[!htbp]
\begin{center}
\caption{Mass values for $\eta_c$ and $\chi_{c0}$ on Ens.~{$B_1$} and Ens.~{$C_1$} respectively.
The last line cites the corresponding result from PDG~\cite{PhysRevD.98.030001}.}
\begin{tabular}
{cccc}\hline\hline
                                   &                 &$\eta_c$         &$\chi_{c0}$         \\
\hline
Ens.~{$B_1$}:                      &Mass[MeV]        &2978(1)        &3454(6)              \\
\hline
Ens.~{$C_1$}:                      &Mass[MeV]        &2970(1)        &3416(13)              \\
\hline
PDG:                               &Mass[MeV]        &2983.9(5)       &3414.71(30)            \\
\hline\hline
\label{tab:mass}
\end{tabular}
\end{center}
\end{table}

In principle, glueball states with the same quantum numbers are also present in the similar energy
range~\cite{Sun:2017ipk}. However, in this lattice calculation,
we have only utilized the quark bilinear operators for the charmonium states and have not observed the sign of the glueballs.

\subsection{Renormalization factor $Z_V$ of electromagnetic current operators}
\label{Renormalization factor}
\begin{figure*}[!htbp]
\begin{minipage}{0.48\linewidth}
  {\includegraphics[width=9cm]{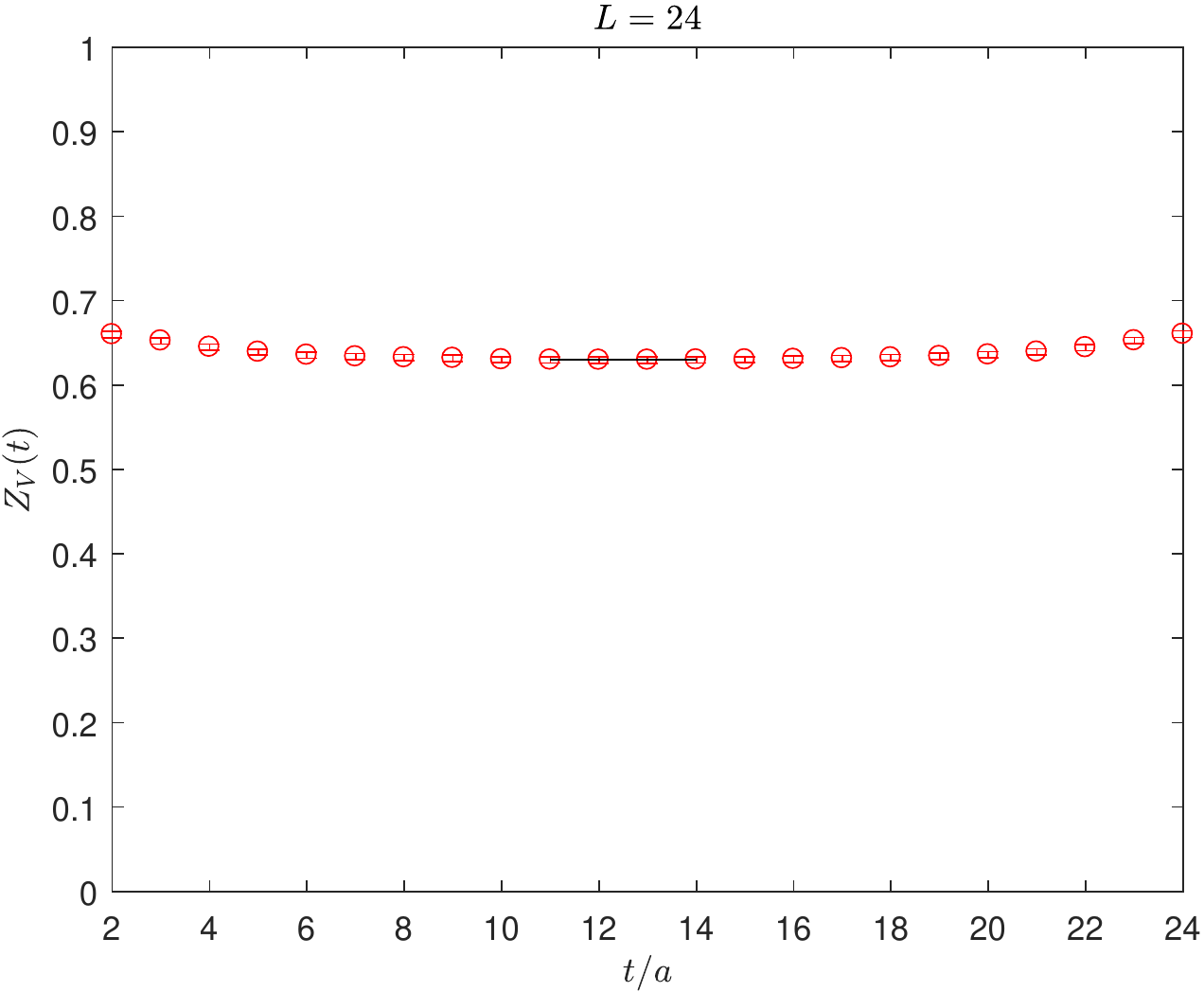}}
  \center (a) Ens.~$B_1$
\end{minipage}
\hfill
\begin{minipage}{0.48\linewidth}
  {\includegraphics[width=9cm]{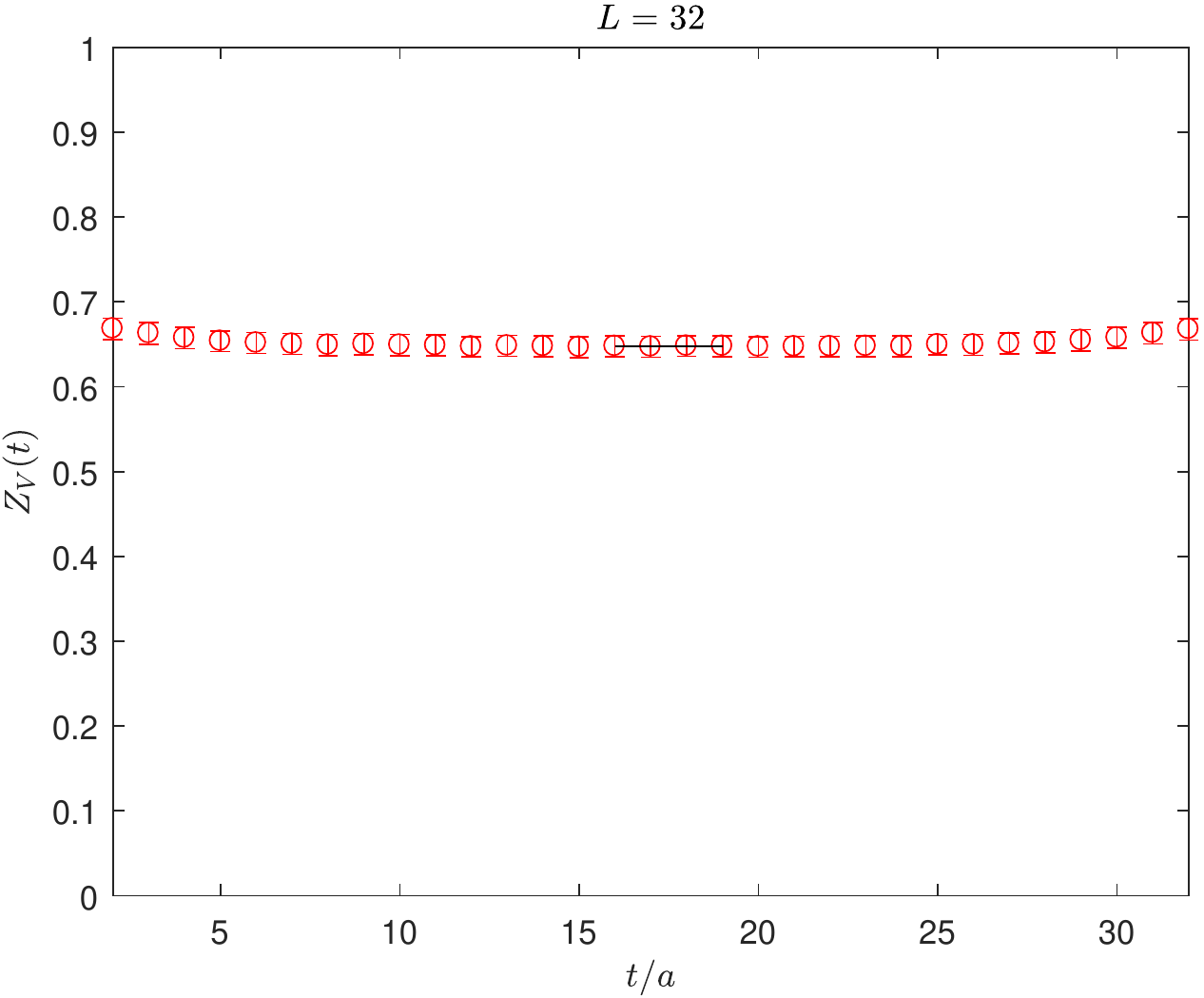}}
  \center (b) Ens.~$C_1$
\end{minipage}
\caption{Renormalization factors $Z_V$ for Ens.~{$B_1$} and Ens.~{$C_1$}.
 The horizontal line segments denote the corresponding values for $Z_V$.}
\label{fig:renormalization factor}
\end{figure*}

The current operators in Eq.~(\ref{eq:three point1}) are electromagnetic
current operators. In principle, they contain all flavors of quarks weighted by the corresponding charges.
Light quark flavors will only enter the question via disconnected diagrams which are neglected in this study.
Only considering the charm quark, we only need to consider the current $\bar{c}\gamma_\rho c(x)$.
A subtlety in the lattice computation is that, with $c(x)/\bar{c}(x)$ being the bare charm/anti-charm quark field on the lattice, composite operators such as the current $j_\rho(x)=Z_V\bar{c}(x)\gamma_{\rho}c(x)$
needs an extra multiplicative renormalization factor $Z_V$ which can be extracted by the ratio
 of two-point function with respect to the three-point function for $\eta_c$~\cite{Dudek:2006ej}:
\beq
Z_V^{\mu}&=&\frac{p^{\mu}}{E(p)}\frac{(1/2)\Gamma_{\eta_c}^{(2)}}{\Gamma^{(3)}_{\eta_c}},
\label{eq:renormalize}
\eeq
where $\mu$ is the Dirac index and we take it to be zero. $\Gamma_{\eta_c}^{(2)}$ and $\Gamma^{(3)}_{\eta_c}$ are two point
correlation function and three point correlation function relevant with $\eta_c$. The explicit forms are:
\beq
\Gamma_{\eta_c}^{(2)}&=&\sum_{\bx}e^{-i\bp\cdot\bx}\langle\mathcal{O}_{\eta_c}(\bx, t)\mathcal{O}_{\eta_c}^{\dagger}(0, 0)\rangle
\\
\Gamma_{\eta_c}^{(3)}&=&\sum_{\bx,\by}e^{-i\bp_f\cdot\bx}e^{i\bq\cdot\by}\langle\mathcal{O}_{\eta_c}(\bx, t_f)\bar{c}\gamma_0{c}(\by,t)\mathcal{O}_{\eta_c}^{\dagger}(0, 0)\rangle\nonumber\\
\eeq
with $\mathcal{O}^\dagger_{\eta_c}$ and $\mathcal{O}_{\eta_c}$ creating and annihilating a state with the quantum number of $\eta_c$ meson, respectively.
Actually, we use the simple local operator,i.e.$\bar{c}\gamma_5{c}$.
According to Eq.~(\ref{eq:renormalize}), we can obtain the multiplicative renormalization factor $Z_V$, and show it in Fig.~\ref{fig:renormalization factor}. The values of the renormalization factor $Z_V$ are $0.6296(18)$ and $0.6476(61)$ for Ens.~$B_1$ and Ens.~$C_1$, respectively.

\subsection{Form factors for two photon decay of $\eta_c$ and $\chi_{c0}$}
\label{sec:formfactory}
\begin{figure*}[!htbp]
\begin{minipage}{0.48\linewidth}
  {\includegraphics[width=9cm]{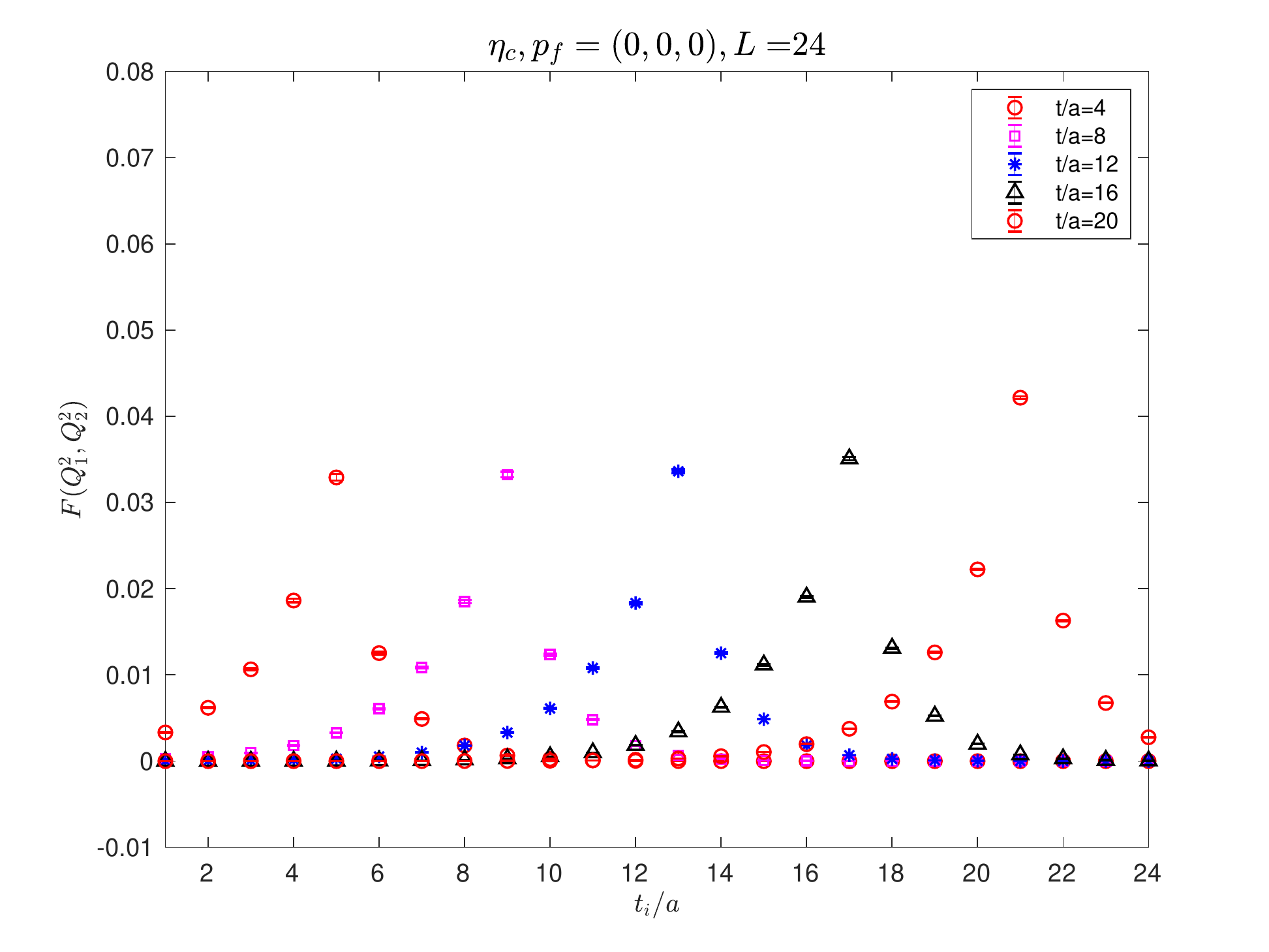}}
  \centerline{$Q_1^2=0GeV^2$}
 \center\ \ \ \ \ \ \ \ \ \ \ \ \ \ \
  \center (a) Ens.~$B_1$
\end{minipage}
\hfill
\begin{minipage}{0.48\linewidth}
  {\includegraphics[width=9cm]{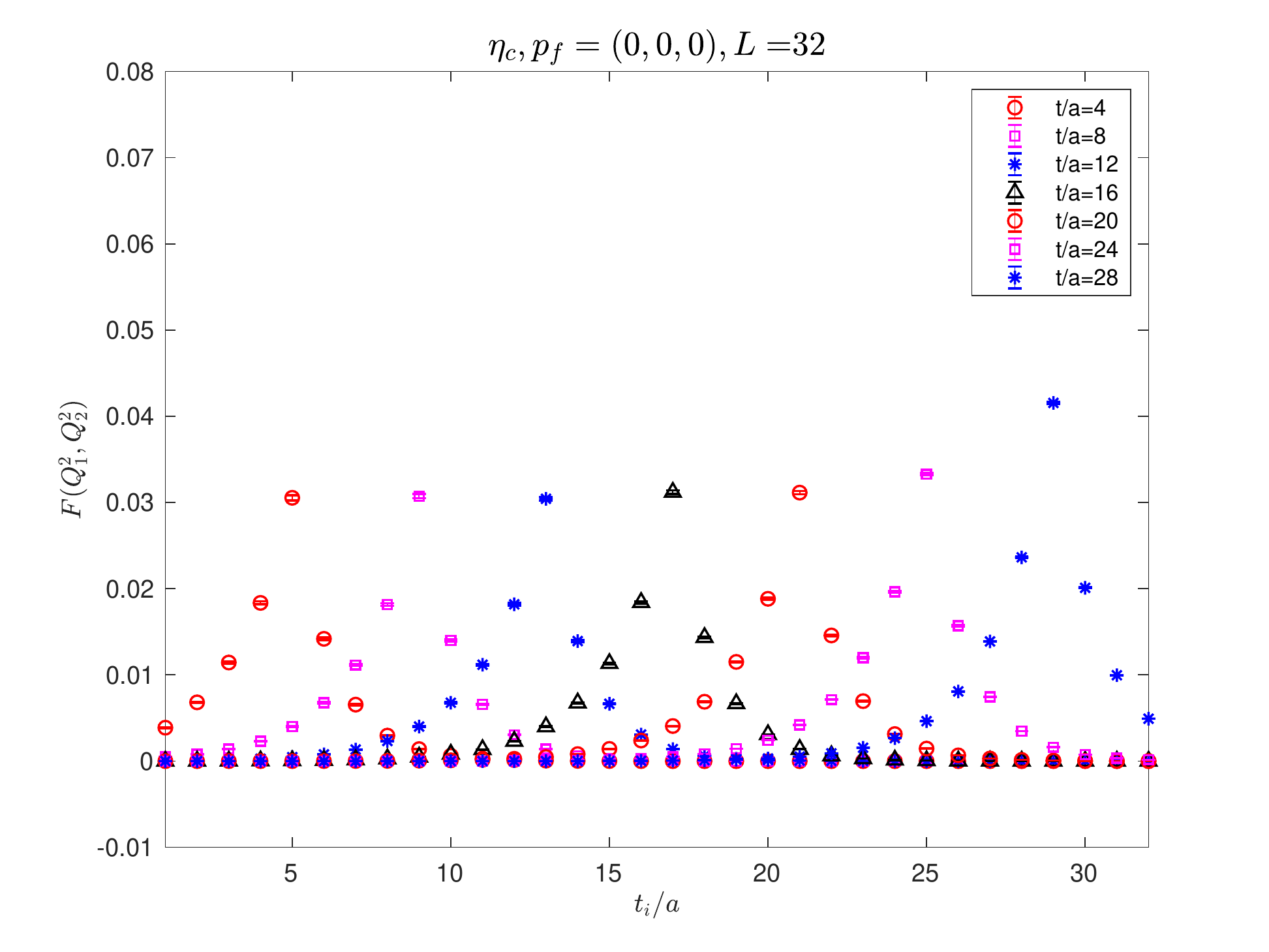}}
  \centerline{$Q_1^2=0GeV^2$}
 \center\ \ \ \ \ \ \ \ \ \ \ \ \ \ \
  \center (b) Ens.~$C_1$
\end{minipage}
\caption{The integrand for insertion positions obtained from simulations on Ens.~$B_1$ (left figure),
 and Ens.~$C_1$ (right figure) respectively. We take $n_2=(-1 -1 -2)$; $n_f=(0\ 0\ 0)$ as an example.
 The insertion positions for lattice size: $24^3\times48$ and lattice size: $32^3\times64$ are $t=4,~8,~12,~16,~20$ and $t=4,~8,~12,~16,~20,~24,~28$ respectively.}
\label{fig:FormFactor_fit_choose_ty}
\end{figure*}

\begin{figure*}[!htbp]
\begin{minipage}{0.48\linewidth}
{\includegraphics[width=9cm]{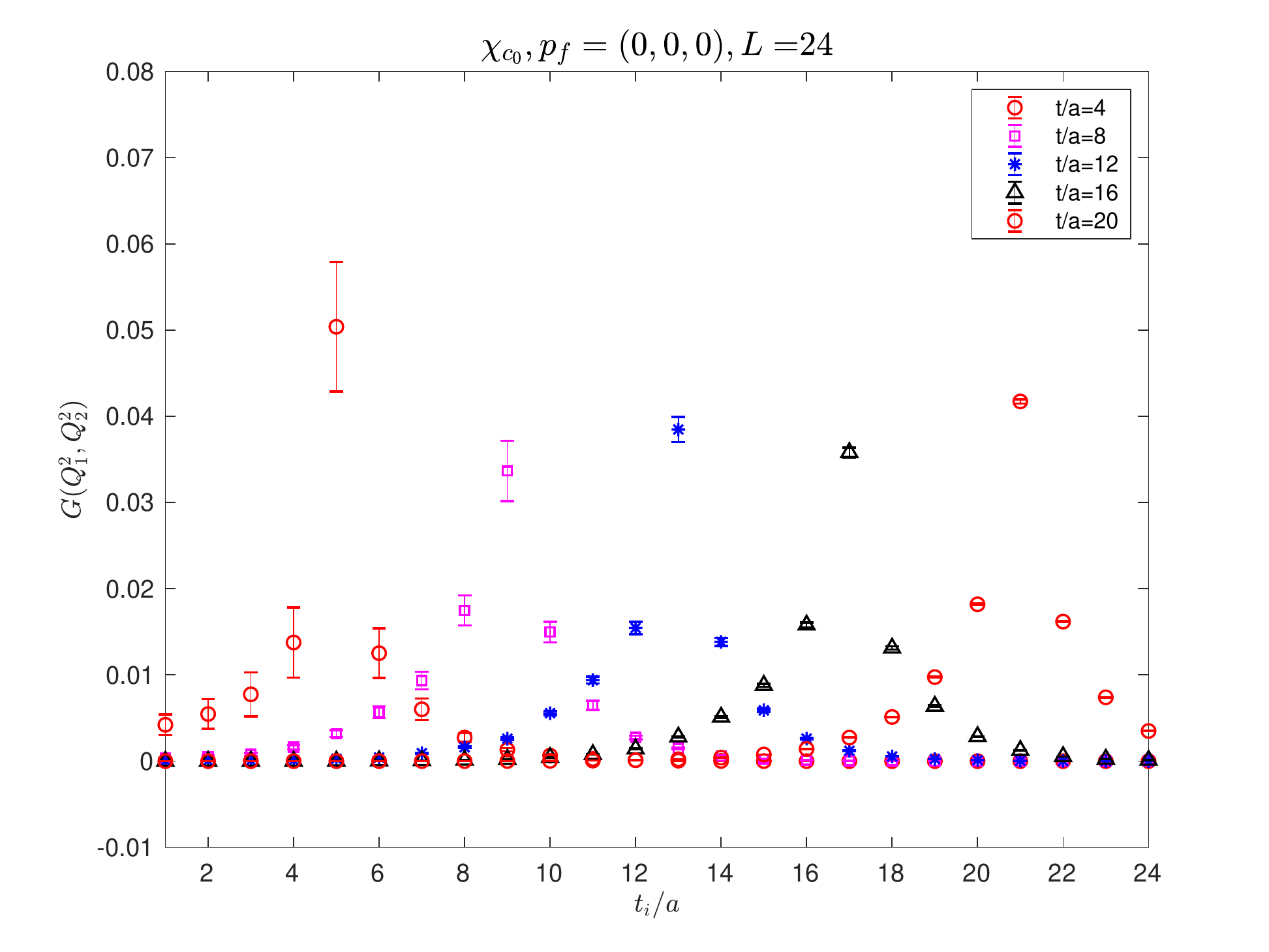}}
\centerline{$Q_1^2=0GeV^2$}
\center\ \ \ \ \ \ \ \ \ \ \ \ \ \ \
\center (a) Ens.~$B_1$
\end{minipage}
\hfill
\begin{minipage}{0.48\linewidth}
{\includegraphics[width=9cm]{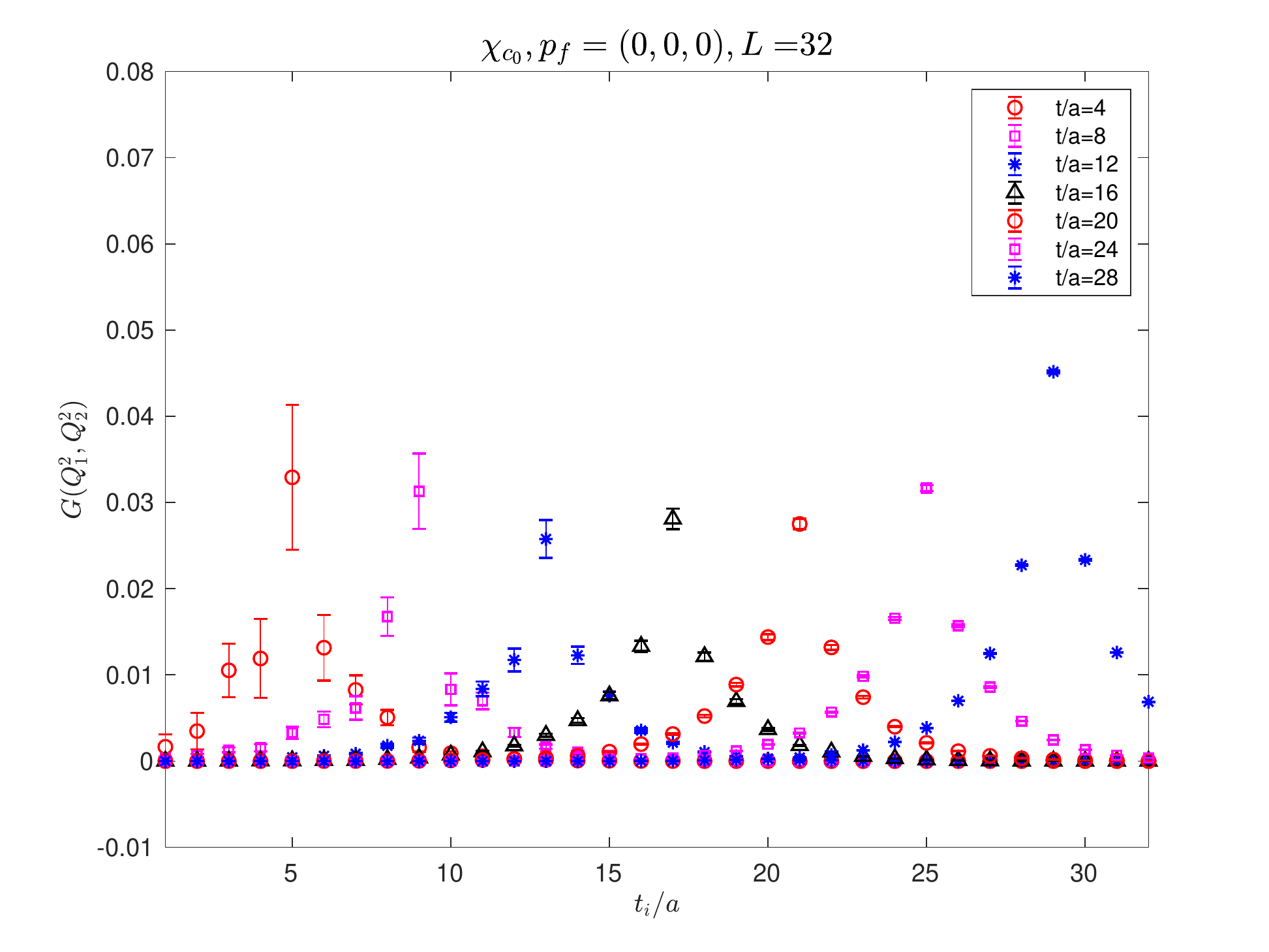}}
\centerline{$Q_1^2=0GeV^2$}
\center\ \ \ \ \ \ \ \ \ \ \ \ \ \ \
\center (b) Ens.~$C_1$
\end{minipage}
\caption{The integrand for insertion positions obtained from simulations on Ens.~$B_1$ (left figure),
and Ens.~$C_1$ (right figure) respectively. We take $n_2=(0 0 -3)$; $n_f=(0\ 0\ 0)$ as an example. The insertion positions for lattice size: $24^3\times48$ and lattice size: $32^3\times64$ are $t=4,~8,~12,~16,~20$ and $t=4,~8,~12,~16,~20,~24,~28$ respectively.}
\label{fig:FormFactor_fit_choose_t}
\end{figure*}

To compute the relevant matrix element in
Eq.~(\ref{master}), we place the meson at a
fixed sink position $t_f$ which is chosen to be $24$  for  Ens.~$B_1$ and $32$ for Ens.~$C_1$, respectively.
These sinks are then used as a sequential source for a backward propagator inversion.
This allows us to investigate  all possible source positions $t_i$. We can then
freely vary the values of $\omega_1$,  $Q^2_1$ and inspect the integrand as a function of $t_i$
in Eq.~(\ref{master}) for a given insertion position $t$.
As an example, in Fig.~\ref{fig:FormFactor_fit_choose_ty} and Fig.~\ref{fig:FormFactor_fit_choose_t}, we show the integrand for the
insertion positions $t=4,8,12,16,20$ for Ens.~$B_1$ and $t=4,8,12,16,20,24,28$ for Ens.~$C_1$
with a particular value $\vec{p}_f=(000)$ for $\eta_c$ and $\chi_{c0}$.

\begin{figure*}[!htbp]
\begin{minipage}{0.45\linewidth}
  \centerline{\includegraphics[width=9.0cm]{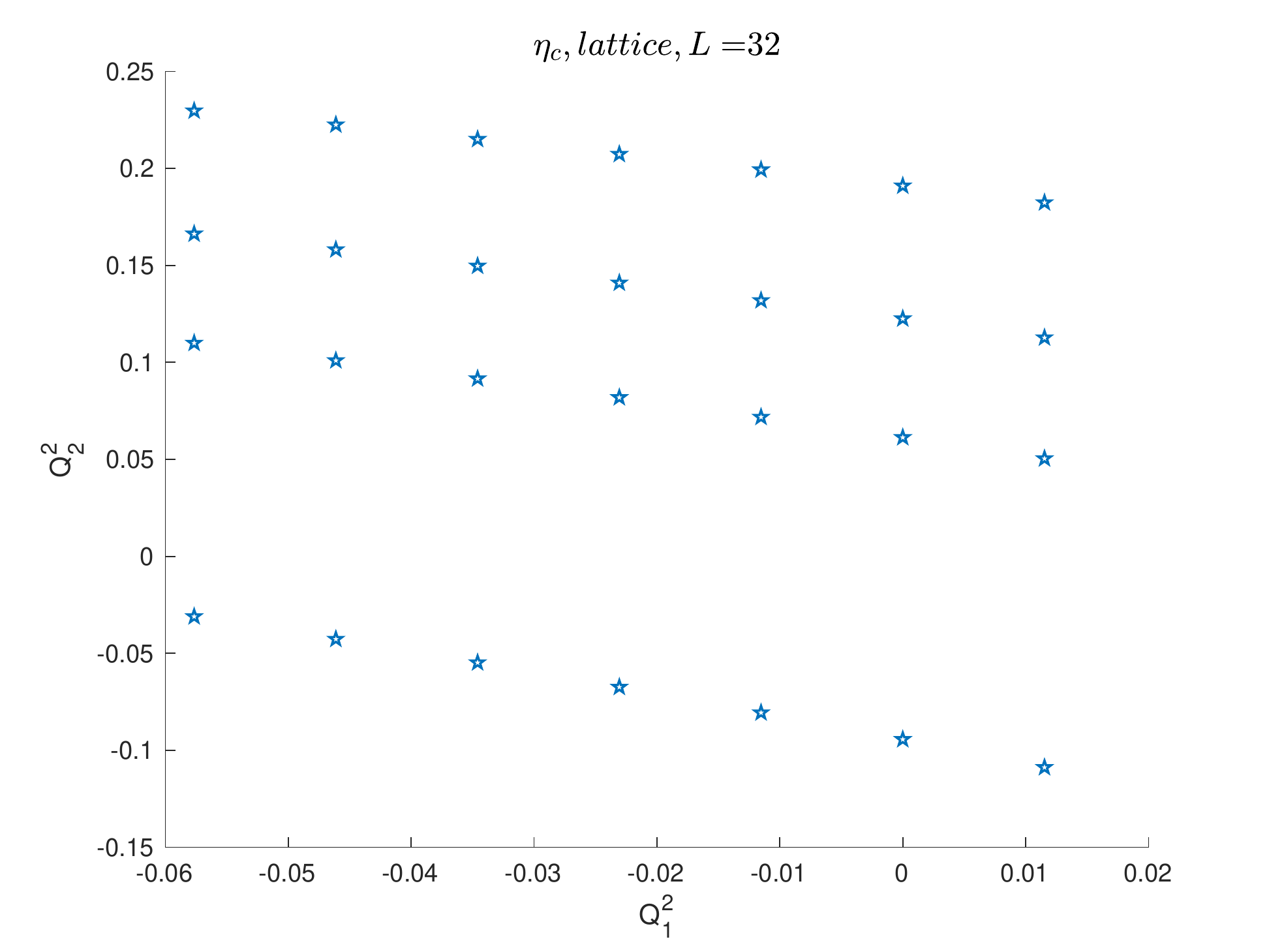}}
  \centerline{Ens.~$C_1$}
  \end{minipage}
\hfill
\begin{minipage}{0.45\linewidth}
  \centerline{\includegraphics[width=8.8cm]{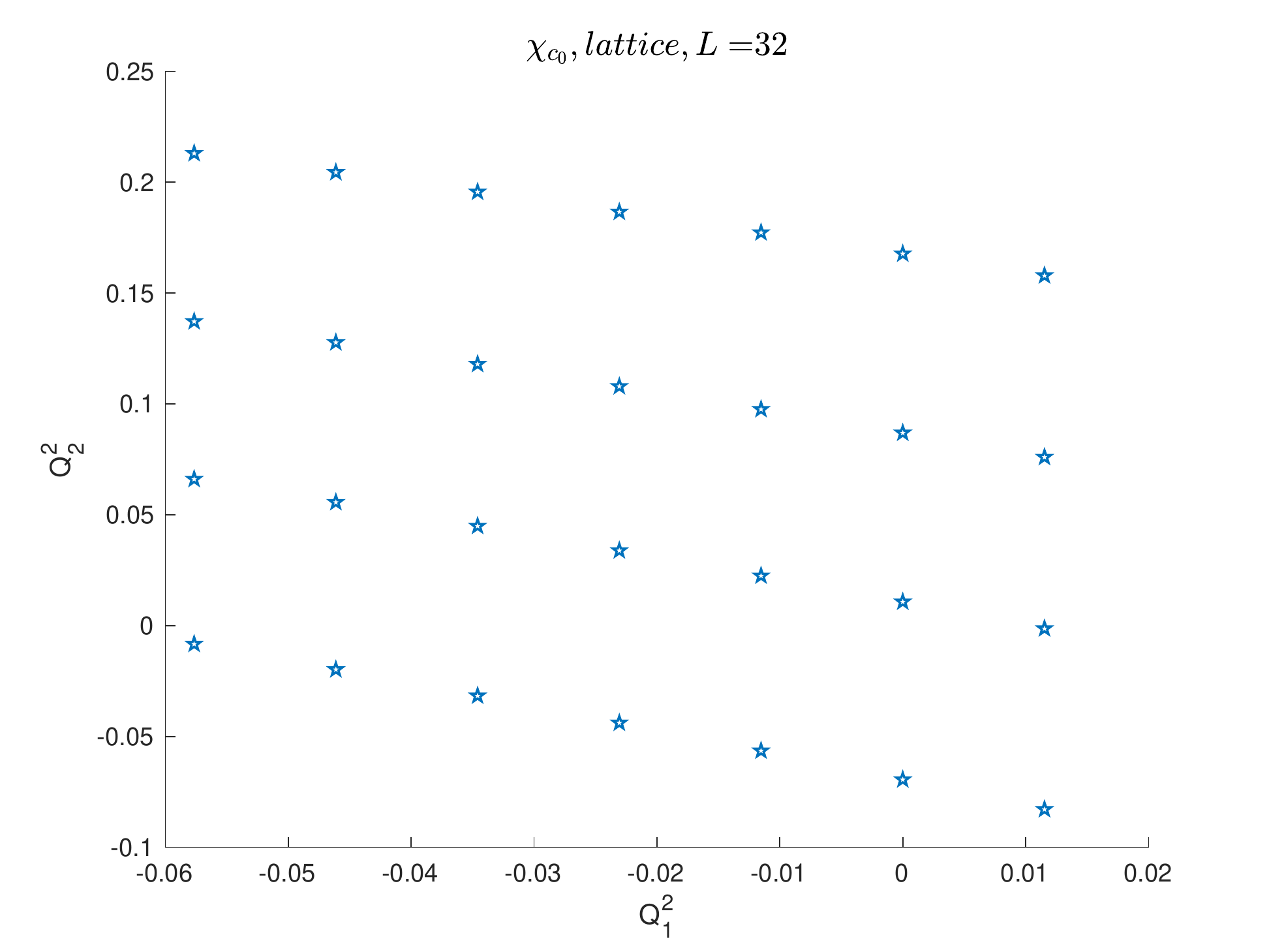}}
  \centerline{Ens.$C_1$}
  \end{minipage}
\caption{The distribution of virtualities $(Q^2_1,Q^2_2)$  (lattice units) for Ens.$C_1$
computed for $\eta_c$ (left panel) and $\chi_{c0}$ (right panel). }
\label{fig:distributions}
\end{figure*}

The computation has to cover the physical interesting kinematic regions.
For this purpose, we have to scan the corresponding parameter space
of the two virtualities $Q^2_1$ and $Q^2_2$.
Basically, we follow the following strategy: we first fix the four-momentum
of $\eta_c$ and $\chi_{c0}$, $p_f=(E,\bp_f)$ and place it on a given time-slice $t_f=T/2$.
In this simulation, we only compute the case of $\bp_f=(0,0,0)$,
and $E$ is simply the mass of $\eta_c$ or $\chi_{c0}$ meson.
Then, we judiciously choose several values
of virtuality $Q^2_1$ around the physical point $Q^2_1=0$.
To be specific, we picked the range $Q^2_1\in [-0.5,+0.1]$~GeV$^2$ on Ens.~$B_1$ and $Q^2_1\in [-0.5,+0.1]$~GeV$^2$ on Ens.~$C_1$,
which satisfies the constraint $Q^2_1>-m^2_\rho$~\cite{Chen:2016yau}.
For a given $\bp_f$, a choice of $\bq_1$ completely specifies $\bq_2$ due to $\bp_f=\bq_1+\bq_2$.
Therefore, we take several choices of $\bq_1=\bn_1(2\pi/L)$ by
changing three-dimensional integer $\bn_1$.
Then the energy of the first photon is also obtained using either the
continuum or the lattice dispersion relations:
\beq
\label{eq:dispersion_continuum}
\omega^2_1 &=&\bq^2_1-Q^2_1\;,
\\
\label{eq:dispersion_lattice}
4\sinh^2(\omega_1/2) &=& 4\sum_i\sin^2(\bq_{1i}/2)-\hat{Q}^2_1\;,
\eeq
 where $\hat{Q}^2_1=4\sinh^2(Q_1/2)$ is the lattice version for the virtuality.
It turns out that we can also compute the virtuality of the second photon,
 since both $\omega_2$ and $\bq_2$ are constrained by the energy-momentum conservation.
One has to make sure that, the values of $Q^2_2$ thus computed
do satisfy the constraint $Q^2_2>-m^2_\rho$ otherwise it is omitted.
This procedure is summarized as follows:
\begin{enumerate}
\item Judiciously choose several values of $Q^2_1$ in a suitable range. We picked 7 values
of $Q^2_1$;
\item Pick different values of $\bn_1$ such that $\bq_1=\bn_1(2\pi/L)$.
As described above, this fixes both $\omega_1$ and $Q^2_2$ using energy-momentum conservation.
 This is done using either the continuum or the lattice dispersion relations.
 To be specific, for each $Q^2_1$, we picked 4 different $\bq_1$;
\item Make sure all values of $Q^2_1,Q^2_2 >-m^2_\rho$, otherwise the choice is simply ignored;
\item For each validated choice above, compute the three-point functions~(\ref{eq:three point1}) and obtain the
hadronic matrix element using Eq.~(\ref{master}).
 \end{enumerate}
 In such a way, we have obtained altogether 28 points on the $(Q^2_1,Q^2_2)$ plane
 around the origin. As an example, the distribution of these virtualities for
 the two mesons are shown in Fig.~\ref{fig:distributions} for the case of lattice
 dispersion relations. One could also do the same thing using the conventional continuum
 dispersion relations. The difference of these two treatments finally will provide us with an estimate
 for the finite lattice spacing error of the calculation.


\begin{figure*}[!htbp]
\begin{minipage}{0.45\linewidth}
  \centerline{\includegraphics[width=9.0cm]{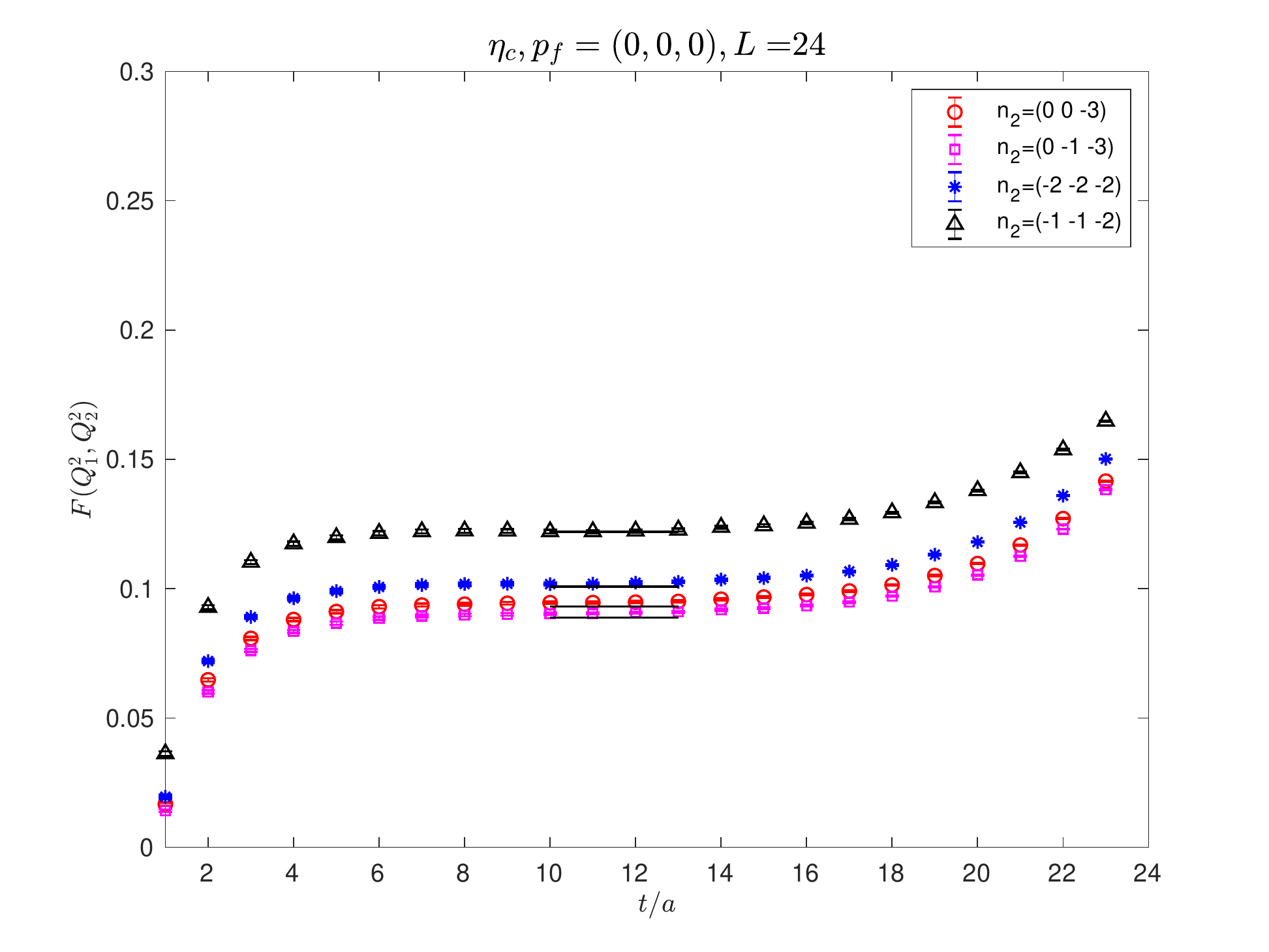}}
  \centerline{Ens.~$B_1$}
  \end{minipage}
\hfill
\begin{minipage}{0.45\linewidth}
  \centerline{\includegraphics[width=8.8cm]{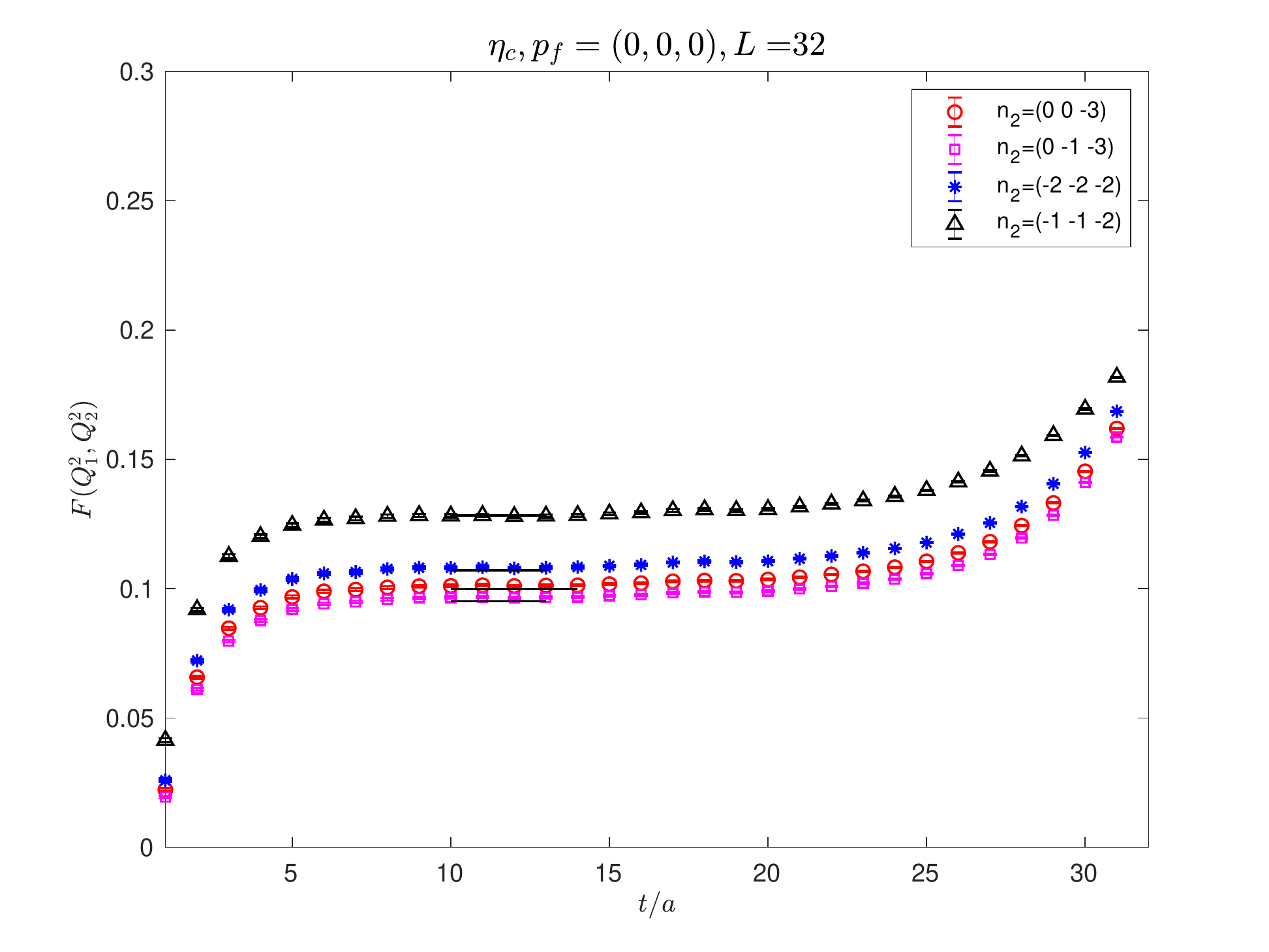}}
  \centerline{Ens.$C_1$}
  \end{minipage}
\caption{The plateaus of the pseudoscalar form factors obtained from integration of $t_i$ for three point function $G_{\mu\nu}(t_i,t)$ with Ens.$B_1$ (left figure), and Ens.$C_1$ (right figure) respectively. We take $Q_1^2=0~GeV^2$; $n_f=(0\ 0\ 0)$ as an example. The errors in these figures are estimated using the conventional jack-knife method.}
\label{fig:FormFactor_plateauy}
\end{figure*}

\begin{figure*}[!htbp]
\begin{minipage}{0.45\linewidth}
\centerline{\includegraphics[width=8.8cm]{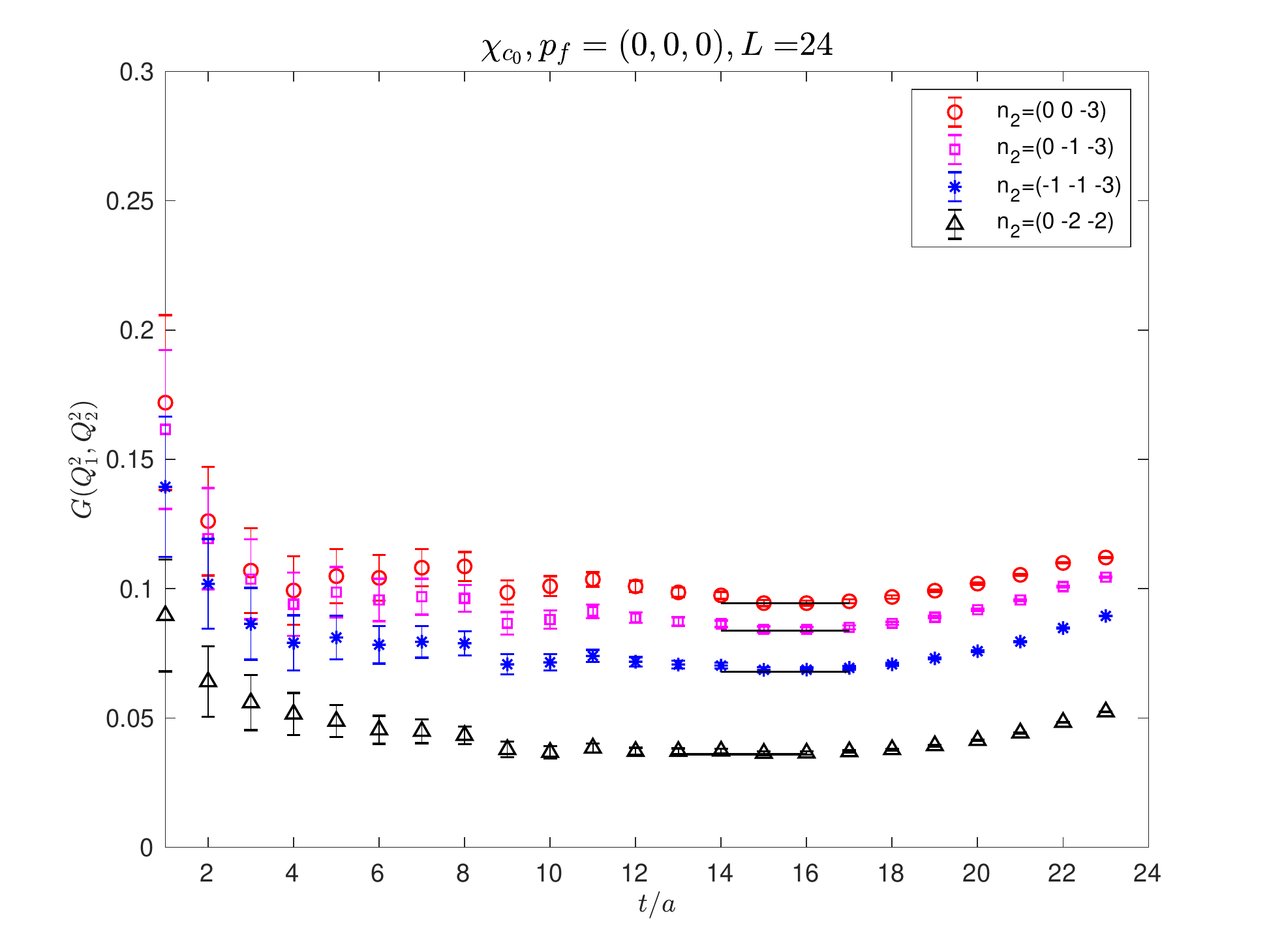}}
\centerline{Ens.~$B_1$}
\end{minipage}
\hfill
\begin{minipage}{0.45\linewidth}
\centerline{\includegraphics[width=8.8cm]{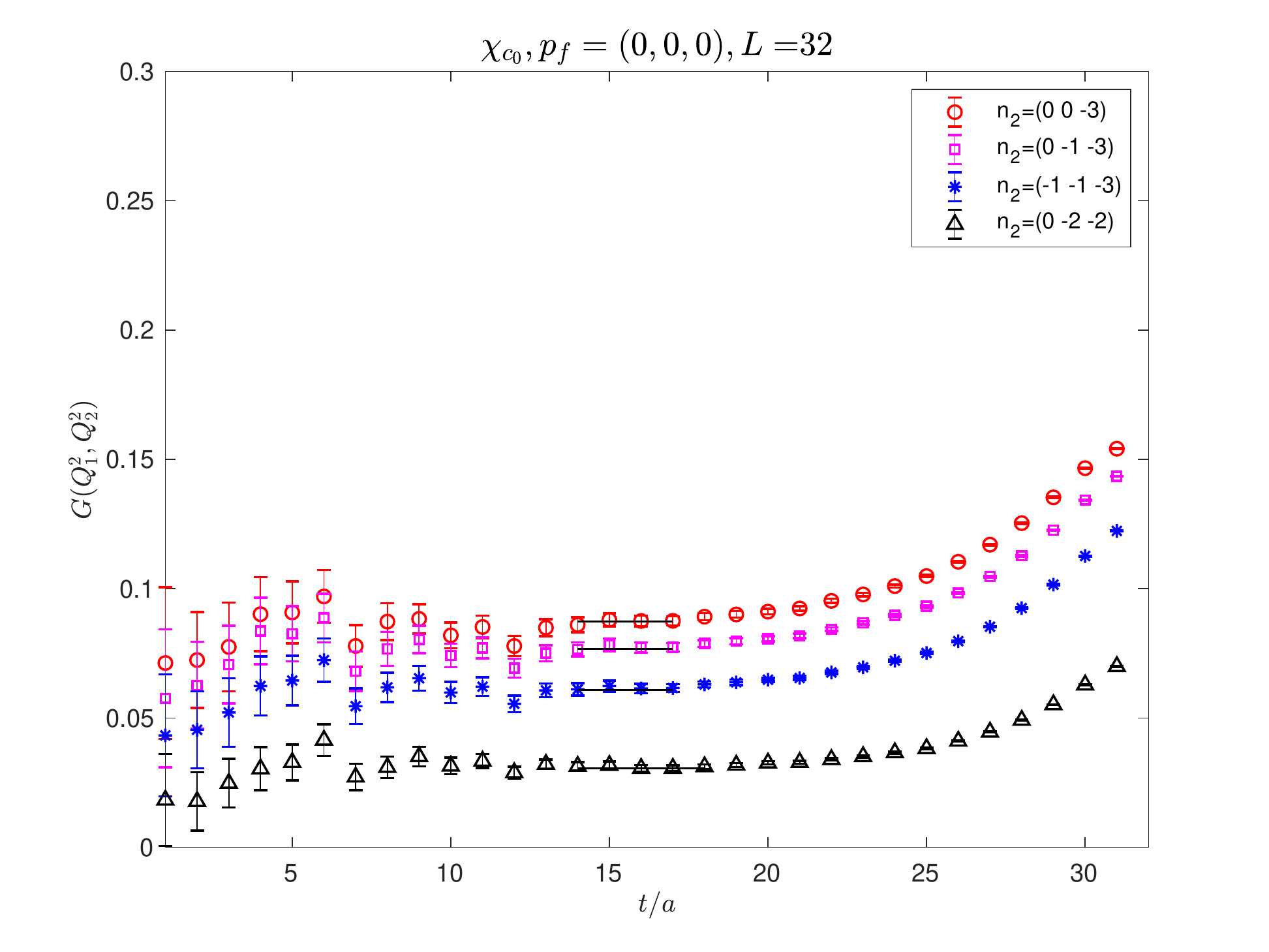}}
\centerline{Ens.$C_1$}
\end{minipage}
\caption{The plateaus of scalar form factor obtained from integration of varies $t_i$ for three point function $G_{\mu\nu}(t_i,t)$ with Ens.$B_1$ (left figure) and Ens.$C_1$ (right figure) respectively. We take $Q_1^2=0~GeV^2$; $n_f=(0\ 0\ 0)$ as an example. The errors in these figures are estimated using the conventional jack-knife method.}
\label{fig:FormFactor_plateau}
\end{figure*}

In our real lattice QCD computation, the integration of $t_i$ in Eq.~(\ref{master}) are
replaced by discrete summation over $t_i$ using trapezoid rule.
The resulting values exhibit a plateau behavior with respect to $t$ which is then utilized
to extract the corresponding form factor. In such a way, we have obtained numerical results
for $F(Q^2_1,Q^2_2)$ and $G(Q^2_1,Q^2_2)$ at 28 different points in the plane of
the two virtualities. As an example, the form factor plateaus for $\eta_c$ are
illustrated in Fig.~\ref{fig:FormFactor_plateauy} for the case of $Q_1^2=0~GeV^2$.
The corresponding case for $\chi_{c0}$ are shown in Fig.~\ref{fig:FormFactor_plateau}.

\subsection{Fitting of the form factor and the physical decay widths}
\label{sec:formfactor_fit}
To obtain the physical decay width, we only need the values of the form factors
at the physical photon point, namely $Q^2_1=Q^2_2=0$.
As we have seen in Fig.~\ref{fig:distributions} the distribution of 28 data points in the $(Q^2_1,Q^2_2)$ plane,
 we could implement cuts in the plane.
For a given value of $Q^2_{\rm cut}>0$, we select the points that satisfy the following inequality:
\be
\sqrt{(Q^2_1)^2+(Q^2_2)^2} \le Q^2_{\rm cut}\;.
\ee
 Obviously, taking a large enough $Q^2_{\rm cut}$ will allow all of the points
 be considered while selecting a smaller value for $Q^2_{\rm cut}$ will take into
 account only those points whose distances are closer than $Q^2_{\rm cut}$.
 On the other hand, for a given value of $Q^2_{\rm cut}$, we could utilize
 different fitting formula to obtain the corresponding values of the form factor.
 Since it is the value at the origin that is directly related to the decay width,
 it is natural to use a polynomial type of fit in both $Q^2_1$ and $Q^2_2$.
 What is more, due to boson symmetry, this function needs to be symmetric with respect
 to $Q^2_1$ and $Q^2_2$.  Therefore, by varying the cut value $Q^2_{\rm cut}$ and various orders
 of polynomials in the virtualities, we could investigate the values
 of the form factors at the physical point.

 To be specific, we adopt a polynomial ansatz up to $Q^2_1$ and $Q^2_2$ to the third power to
 fit the data of the form factor. For $\eta_c$ meson, we use:
\beq
\label{eq:formfactor_fit_3y}
F(Q_1^2,Q_2^2)&=&a_0+a_1(Q_1^2+Q_2^2)\nonumber \\
&+&a_2(Q_1^4+Q_2^4)+a_3Q_1^2Q_2^2\nonumber \\
&+&a_4(Q_1^6+Q_2^6)+a_5(Q_1^2Q_2^4+Q_2^2Q_1^4)\;.
\eeq
 and a similar form for the $\chi_{c0}$ form factor $G(Q^2_1,Q^2_2)$.
 Note that $a_0\equiv F(0,0)$ is the form factor at the physical photon point which
 is directly related to the decay width of the meson.
 Polynomial forms with less terms, i.e. with only up to first or second powers in $Q^2_1$ and $Q^2_2$. have also
 been attempted. Note that this implies that we are fitting the data points with $2$, $4$ and $6$ parameters, respectively
 since terms at the same orders of $Q^2_1$ and $Q^2_2$ should be included or excluded on the same footing.
 In all cases, correlated fits are performed.
 Depending on the number of points taken into account which is controlled by $Q^2_{\rm cut}$,
 and the number of terms kept in the fitting polynomial,
 we finally arrive at the values for the form factors at the origin,
 namely $F(0,0)=a_0$ for both ensembles.
 Similar procedures have been implemented as well for $\chi_{c0}$,
 resulting in the values for $G(0,0)$.

 The  fitting procedures described above can be implemented
 using either the continuum or lattice version of the dispersion
 relations as indicated in Eq.~(\ref{eq:dispersion_continuum}) or Eq.~(\ref{eq:dispersion_lattice}).
 The procedure can be carried out for either pseudo-scalar or scalar meson on either of the
 two ensembles utilized in this calculation. Therefore, we carry out the fitting procedure in 8 different cases.
 The difference between the corresponding results obtained from different dispersion relations
 will then inform us about lattice artifacts of the calculation.

\begin{figure*}[!htbp]
\begin{minipage}{0.48\linewidth}
  {\includegraphics[width=9cm]{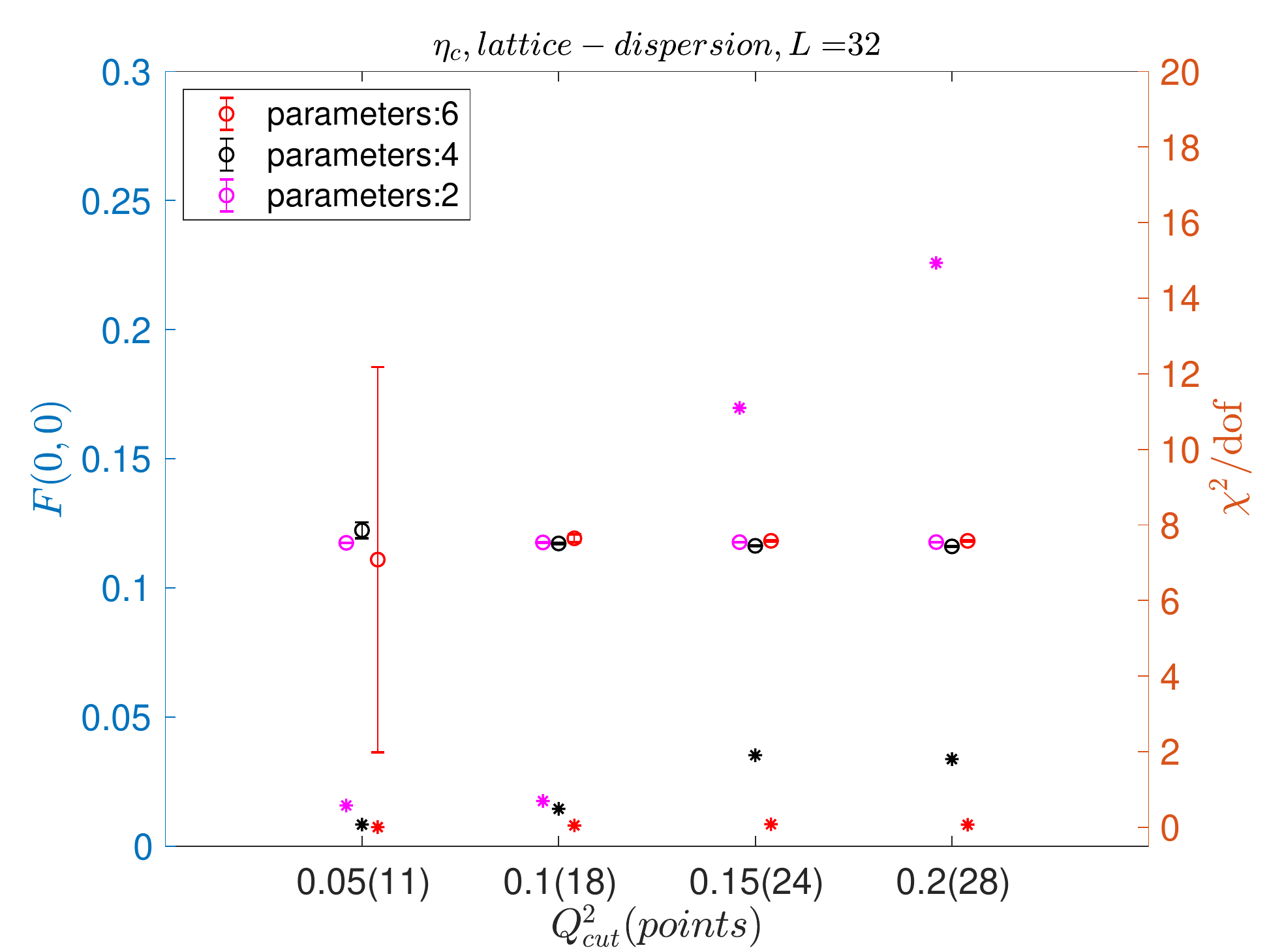}}
 \center\ \ \ \ \ \ \ \ \ \ \ \ \ \ \
 \center (a) Ens.~$C_1$ ~ $\eta_c$
\end{minipage}
\hfill
\begin{minipage}{0.48\linewidth}
  {\includegraphics[width=9cm]{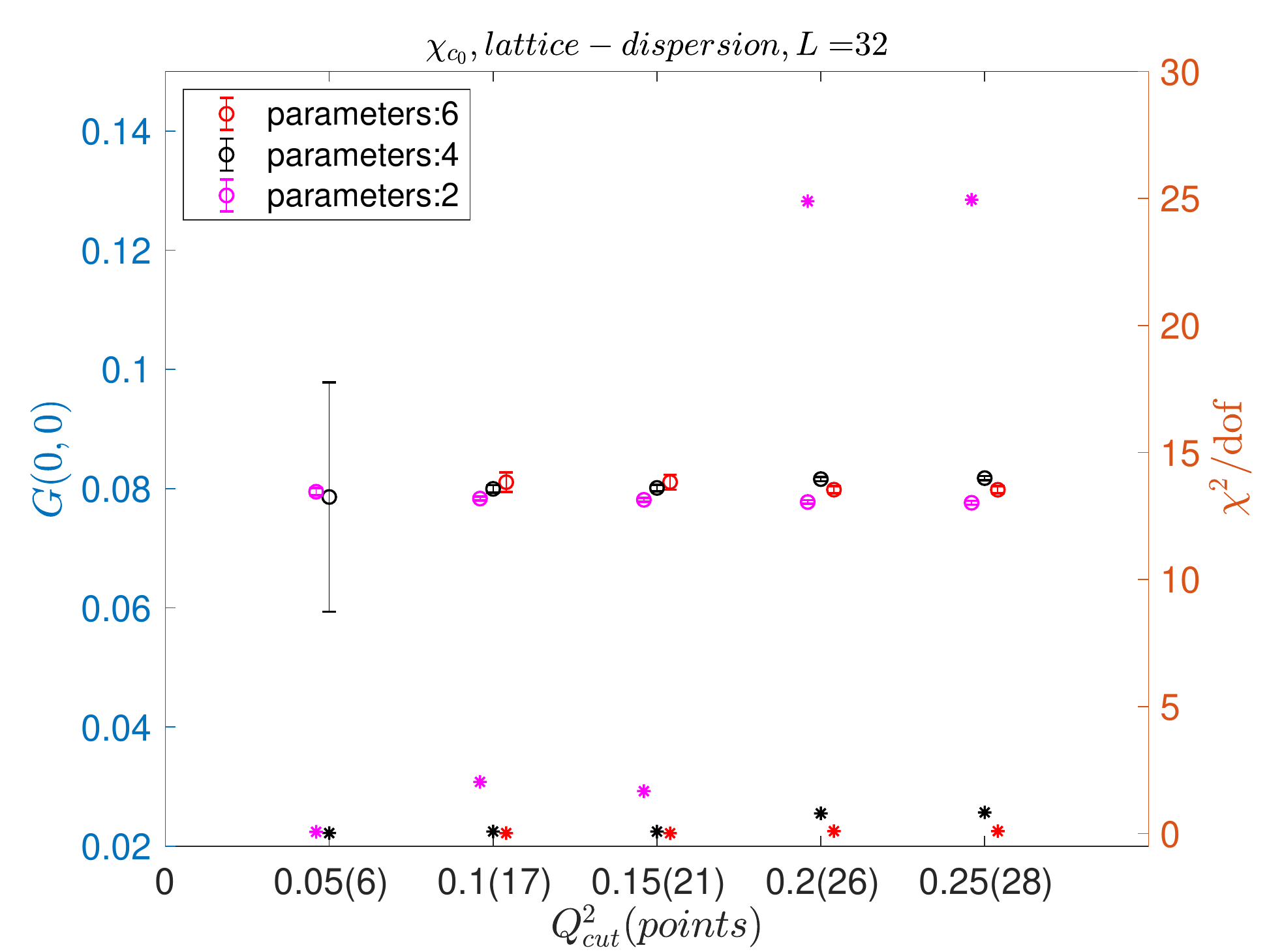}}
 \center\ \ \ \ \ \ \ \ \ \ \ \ \ \ \
  \center (b) Ens.~$C_1$ ~ $\chi_{c0}$
\end{minipage}
\caption{The fiting results for $\eta_c$ and $\chi_{c0}$ on Ens.C1 using lattice dispersion relations. The horizontal axis denotes the cut values $Q^2_{\rm cut}$ while the vertical axis
 indicates the values for $F(0,0)$ or $G(0,0)$ together with the errors (data points with error-bars). The integers in brackets along the horizontal axis indicate the number of data points below $Q^2_{\rm cut}$. Points without errors show the corresponding values of $\chi^2$/dof, the values of which is obtained to the right edge of each box.}
\label{fig:cut_latt_L32}
\end{figure*}

 As an illustration, in Fig.~\ref{fig:cut_latt_L32}, the fitting results for $\eta_c$ and $\chi_{c0}$ on Ens.C1 using
 lattice dispersion relations are shown.
 Here the horizontal axis denotes the cut values $Q^2_{\rm cut}$ while the vertical axis
 indicates the values for $F(0,0)$ or $G(0,0)$ together with the errors (data points with error-bars).
 For each fixed value of $Q^2_{\rm cut}$, we have performed three fits with $2$, $4$ and $6$ parameters.
 These three points are slightly shifted horizontally so that they are recognizable.
 The corresponding values of $\chi^2/\mbox{dof}$ for each fit are also shown as points without error-bars.
 By inspecting the plot, we get a feeling about the consistency and quality of these different fits and
 the differences among the values of $F(0,0)$ can also offer us an estimate of the systematics for the fitting procedure.

 Having obtained  these 8 plots, we proceed as follows:
\begin{itemize}
\item For each of these plots, say the example above,  we pick the case with lowest value of $\chi^2/\mbox{dof}$ as
the final result for $F(0,0)$ together its statistical error in this particular case.

\item We further attribute a systematic error arising from the fitting procedure by taking the largest
difference in the central values of $F(0,0)$ with comparable $\chi^2/\mbox{dof}$.
 This then yields the result for $F(0,0)$ with a certain type of dispersion relations
 on a particular ensemble.

\item By comparing the difference in $F(0,0)$ between the two different dispersion relations,
 we further assign a systematic error, which is taken to be the difference between the two values,
 arising from the lattice spacing.

\item Similar procedure can be applied to $\chi_{c0}$ as well.
\end{itemize}
In such a way, we have obtained the results of $F(0,0)$ and $G(0,0)$ on two ensembles which
are explicitly listed below:
\beq
F(0,0)_{B1} &=& 0.1283(1)(3)(77)\;,
\\
F(0,0)_{C1} &=& 0.1240(4)(13)(68)\;,
\\
G(0,0)_{B1} &=& 0.1017(7)(102)(126)\;,
\\
G(0,0)_{C1} &=& 0.0907(8)(19)(90)\;,
\eeq
In these expressions, the first error is statistical, the second
is the error from the fitting procedure described above and the third one is the finite lattice error estimated from using two different dispersion relations. It is clearly seen that, in all cases, the results are dominated by systematic errors, especially the finite lattice spacing errors. In fact, two results on two different ensembles are consistent within this estimate of finite lattice spacing errors. We therefore decide not to make any continuum extrapolations. Obviously, computations with more values of lattice spacings will be crucial to nail down these large lattice spacing errors.

To compare with previous lattice computations, we notice that the result for $\Gamma(\eta_c\rightarrow\gamma\gamma)$ is slightly larger than previous result presented in Ref.~\cite{Chen:2016yau} using the same set
of configurations. This difference might come from the fact that the mixing of the $\eta_c$ and $\chi_{c0}$
in the twisted mass setup was not fully disentangled in the previous calculation in Ref.~\cite{Chen:2016yau}.
It is found that, if we were not using properly chosen operator that mix with both
the $\eta_c$-like and $\chi_{c0}$-like interpolating operators, we would not
be able to observe the correct $\chi_{c0}$ signal as we have discussed at
the end of subsec~\ref{simulation configuration}.

Finally, let us convert the results in the form factors into corresponding ones in the decay widths. We simply add all the errors in the form factors in quadrature and neglect the errors in the mass of the mesons. This then leads to the following results for the decay widths:
\beq
\label{eq:calcdata}
\Gamma(\eta_c\rightarrow\gamma\gamma)_{B1} &=& 1.62(19)\;\rm KeV ,\nonumber
 \\
 \Gamma(\eta_c\rightarrow\gamma\gamma)_{C1} &=& 1.51(17)\;\rm KeV ,\nonumber
\\
\Gamma(\chi_{c0}\rightarrow\gamma\gamma)_{B1} &=& 1.18(38)\;\rm KeV ,\nonumber
\\
\Gamma(\chi_{c0}\rightarrow\gamma\gamma)_{C1} &=& 0.93(19)\;\rm KeV ,
\eeq
These are to be compared with the following values given by PDG:
\beq
\label{eq:PDGdata}
\Gamma(\eta_c\rightarrow\gamma\gamma)_{PDG} &=& 5.02(51) \rm KeV ,\nonumber
\\
\Gamma(\chi_{c0}\rightarrow\gamma\gamma)_{PDG} &=& 2.20(22) \rm KeV ,
\eeq
These numbers are all in the same ballpark as the experimental ones though still somewhat smaller. However, since no controlled continuum extrapolations have been performed yet, it is still premature to draw any conclusions for the discrepancy. Our large estimated finite lattice errors offer some hint that this might be the major source of errors.
In the future, more studies at different lattice spacing s are needed in order to control the lattice artifacts in a systematic fashion.
Another source of systematic error could come from the mixing with the nearby glueball states. So far, no lattice calculations have considered such an effect. Considerable efforts are needed in the future in order to bring these into account.

\section{Conclusions}
\label{conclusion}

 In this exploratory study, we calculate the two-photon decay width for $\eta_c$ and $\chi_{c0}$ using
 unquenched $N_f=2$ twisted mass fermions.
 The computation is done with two lattice ensembles (coarser and finer)
 at two different lattice spacings.
 The mass spectrum for the $\eta_c$ and $\chi_{c0}$ meson state are obtained by solving a generalized eigenvalue problem
 which disentangles parity mixing between the two mesons.

 Our results for the decay width $\Gamma(\eta_c\rightarrow\gamma\gamma)$ and $\Gamma(\chi_{c0}\rightarrow\gamma\gamma)$ are summarized in Eq.~(\ref{eq:calcdata}) for two ensembles utilized in this computation. With only two ensembles we only estimate the finite lattice spacing errors for each ensemble and no continuum extrapolations are performed. Albeit without the continuum extrapolations, our results are in the right ballpark as the PDG values shown in Eq.~(\ref{eq:PDGdata}).

 In the future, lattice calculations with more values of lattice spacings are definitely needed so as to control the finite lattice spacing errors which appear to be a dominant source. Meanwhile, the disconnected contributions is a good supplement to this study. Possible mixing with the gluonic excitations should also be studied.
 It's also helpful to research on unquenched configurations with other ensembles or other methods. We also expect more precise experiments on double-photon decays of charmonium.

\section*{Acknowledgments}

 The authors would like to thank the European Twisted Mass Collaboration (ETMC)
 to allow us to use their gauge field configurations. Our thanks also go to
 National Supercomputing Center  in Tianjin (NSCC). This work is supported in part by the National Science Foundation of China (NSFC) under the Project No. 11505132, No. 11335001, No. 11275169, No. 11405178, No. 11575197, No. 11935017, No. 11575196, No. 11875169,
 No. 11775229. It is also supported in part by the DFG and the NSFC (No. 11261130311) through funds provided to the Sino-Germen CRC 110 symmetries and the Emergence of Structure in QCD. This work is also funded in part by National Basic Research Program of China (973 Program) under code number 2015CB856700. M. Gong and Z. Liu are partially supported by the Youth Innovation Promotion Association of CAS (2013013, 2011013). This work is also supported by the Scientific Research Program Funded by Shaanxi Provincial Education Department under the grant No. 19JK0391, and Natural Science Basic Research Plan in Shaanxi Province of China (Program No. 2019JM-001).

 \input two_photon_decay_reply1.bbl

\end{document}